\documentclass[letterpaper]{sig-alternate-2013}

\usepackage{epsfig}
\usepackage{epstopdf}
\usepackage{colortbl}
\usepackage{subfigure}
\usepackage{booktabs}
\usepackage{multirow}
\usepackage[ruled,vlined]{algorithm2e}
\usepackage{algorithm2e}
\newcommand{\mypara}[1]{{\smallskip \bf #1}\hspace{0.1in}}

\begin{document}
%



\title{Trajectory Recovery From Ash: User Privacy Is NOT Preserved in Aggregated Mobility Data}


\numberofauthors{6} 
%
\author{
%
%
\alignauthor
Fengli Xu\\
       \affaddr{Tsinghua University}\\
       \email{xfl15@mails.tsing-\\ hua.edu.cn}
\alignauthor
Zhen Tu\\
       \affaddr{Tsinghua University}\\
       \email{tuz16@mails.tsing-\\ hua.edu.cn}
\alignauthor
Yong Li\\
       \affaddr{Tsinghua University}\\
       \email{liyong07@tsinghua.edu.cn}
\and  
\alignauthor
Pengyu Zhang\\
       \affaddr{Stanford University}\\
       \email{pyzhang@stanford.edu}
\alignauthor
Xiaoming Fu\\
       \affaddr{University of Goettingen}\\
       \email{fu@cs.uni-goettingen.de}
\alignauthor
Depeng Jin\\
       \affaddr{Tsinghua University}\\
       \email{jindp@tsinghua.edu.cn}
}

%
%
%
%

\maketitle
\begin{abstract}

Human mobility data has been ubiquitously collected through cellular networks and mobile applications, and publicly released for academic research and commercial purposes for the last decade. Since releasing individual's mobility records usually gives rise to privacy issues, datasets owners tend to only publish aggregated mobility data, such as the number of users covered by a cellular tower at a specific timestamp, which is believed to be sufficient for preserving users' privacy.
However, in this paper, we argue and prove that even publishing aggregated mobility data could lead to privacy breach in individuals' trajectories. We develop an attack system that is able to exploit the uniqueness and regularity of human mobility to recover individual's trajectories from the aggregated mobility data without any prior knowledge. By conducting experiments on two real-world datasets collected from both mobile application and cellular network, we reveal that the attack system is able to recover users' trajectories with accuracy about 73\%$\sim$91\% at the scale of tens of thousands to hundreds of thousands users, which indicates severe privacy leakage in such datasets. Through the investigation on aggregated mobility data, our work recognizes a novel privacy problem in publishing statistic data, which appeals for immediate attentions from both academy and industry.
\end{abstract}

\keywords{Trajectory privacy; aggregated mobility data; statistic data privacy}

\section{Introduction}

With the prevalence of mobile devices, human mobility data has been ubiquitously collected through cellular networks and mobile applications, and publicly released for academic research and commercial purposes. For example, one of the largest social applications in China, Wechat, has made the real-time population information monitored by the mobile applications available through public interface \cite{wang2015data}. Apple recently updates its privacy policy to compel users to allow for sharing mobility data with its partners and licensees \cite{apple2015polilcy}. One of the world-leading mobile operators, Orange, has made a significant number of cellular accessing records available to researchers since 2013 \cite{blondel2012data}. One major concern of such data release is how to preserve the privacy of mobile users?

To preserve users' privacy, instead of providing each user's trajectory records, the data owners tend to only publish the aggregated mobility data, such as the number of users covered by a cellular tower at a specific timestamp.  For example, French XData project releases a cellular dataset that only reports the population density of each region \cite{acs2014case}. Mobile operators in China share the real-time number of mobile users at specific locations with some automobile and real estate companies \cite{dengta}. Such statistic data of mobile users' mobility is of great utility in numerous applications, such as epidemic controlling, transportation scheduling and business intelligence. More importantly, most of these data providers believe that such statistical mask can preserve users' privacy because the adversaries cannot distinguish each individual's records. In this paper, we demonstrate that this assumption is false since mobile users' trajectories can be recovered with high accuracy from the statistically-masked aggregated mobility dataset.

Releasing statistic data of mobile users' mobility could result in privacy leakage in their trajectories simply because two key characteristics of human mobility. Firstly, a single user's mobility pattern is coherent and regular, which makes their trajectories highly predictable\cite{song2010limits}. For example, we can observe similar trajectories of a single user across different days, i.e., always commuting between office and home during weekdays. Secondly, a user's mobility pattern is significantly different from others, which allows the adversaries to uniquely re-identify his or her trajectory. Aggregated mobility data provides masked mobility records of mobile users in each time slot despite that we cannot identify those belonging to the same individuals, which is referred to as ``ash'' of the original trajectories. However, our experiment reveals that these two key insights make it possible for the adversaries to identify the mobility records created by a single user, which is equivalent to recovering individual's trajectory from ``ash''.

\begin{table*}[t]
\centering
\begin{tabular}{|c|c|c|c|c|c|}
\hline
\scriptsize Mobile Operator & \scriptsize Time & \scriptsize Duration & \scriptsize Location & \scriptsize Data Source& \scriptsize Information Type\\
\hline\hline
\scriptsize AT\&T~\cite{isaacman2011ranges, isaacman2012human} & \scriptsize 2009 & \scriptsize 2-4 months & \scriptsize NY and LA, USA & \scriptsize voice call and message & \scriptsize anonymized individual mobility information\\
\hline
\scriptsize Sprint~\cite{seshadri2008mobile, wang2013inferring} & \scriptsize 2010 & \scriptsize 1 month & \scriptsize Whole USA & \scriptsize voice call records & \scriptsize anonymized individual call information\\
\hline
\scriptsize Orange~\cite{wesolowski2012quantifying, saravanan2013exploring} & \scriptsize Since 2011 & \scriptsize 2-5 months & \scriptsize  Ivory Coast & \scriptsize voice and text message & \scriptsize aggregated mobility information\\
\hline
\scriptsize Tele. Italia~\cite{douglass2015high} & \scriptsize 2013 & \scriptsize 2 months & \scriptsize Milan, Italy & \scriptsize voice, message and data & \scriptsize aggregated mobility information by towers\\
\hline
\scriptsize China Tele.~\cite{wang2015understanding} & \scriptsize 2014 & \scriptsize 1 month & \scriptsize Shanghai, China & \scriptsize data accessing and logs & \scriptsize aggregated traffic information by towers\\
\hline
\end{tabular}
\caption{Examples of the mobility data releasing by operators from different countries.}\label{table:dataset}
\end{table*}
\normalsize

To recover user's trajectory, we have to answer one fundamental question --- are two masked mobility records created by a single user or by different users? Once solving this problem, we can recover users' trajectories easily. We address this problem by leveraging three general but key facts we made from real-world mobility datasets. First, mobile users tend to have low mobility during nighttime because of natural sleeping cycle, i.e., more than 90\% of nighttime mobility records are created in the same location. Therefore, we are able to estimate each mobility record's next location and associate those belonging to same users together by maximizing the likelihood. Through this approach, we can recover the nighttime trajectory of each user. Second, utilizing the fact that a user's mobility is continuous during daytime, we are able to estimate the mobile users' next locations based on their current mobility. Specifically, we use a velocity model to estimate the next location of a user, and then associate unassigned location points with nighttime trajectories by minimizing the estimation error, which expands the nighttime trajectories and recovers the whole trajectories of each day. Third, exploiting the fact that mobile users' mobility trajectories are of high regularity while significantly different from each others, we identify and associate users' trajectories across days by measuring their similarity, which finishes recovering the whole trajectory of each mobile user. It worth noting that the above attack system only exploits the universal characteristics of user's mobility, and hence does not require any prior knowledge or parameters.

By utilizing two large-scale mobility datasets collected from both mobile application and cellular network, we carry out a thorough investigation on whether aggregated mobility data preserves users' privacy or not. We find following two important observations. First, mobile users' privacy is not preserved in aggregated mobility data, i.e., we are able to correctly recover 73\%$\sim$91\% mobility trajectory of each user in a cellular network dataset that contains tens of thousands to hundreds of thousands of mobile users. Second, we evaluate how do the key factors of datasets impact on the privacy leakage. The investigated factors include the spatial and temporal resolution of the released data and the number of users released. Surprisingly, we find out that spatial and temporal resolution has little impacts on privacy preservation, while the attack model is effective even in large-scale mobility data. These results indicate that the attack model is robust to different settings of datasets, and the recognized privacy problem is severe and universal in the aggregated mobility datasets.


The rest of this paper is organized as follows. In Section 2, we identify and define the privacy issue in aggregated mobility data to motivate our study. We introduce two real-world datasets and explore the feasibility to recover individual's trajectory in Section 3, while the design of attack system is discussed in Section 4. With the attack system, we evaluate the privacy leakage on investigated datasets in Section 5. Finally, after discussing the potential mitigation solution and future work in Section 6,
we summarize the related works in Section 7 and conclude our discovery in Section 8.


\section{Motivation}

\subsection{How Operators Release Their Datasets?}

Table \ref{table:dataset} summaries a few examples of mobility datasets released by mobile operators. These datasets have various duration and record mobile users' mobility through the service of voice call, SMS, data plan usage, etc. Operators offered the anonymized individual mobility records in their early release around 2009 and 2010. Then, researchers immediately found out that mobility records generated by each user are significantly different from others, which form \emph{quasi-identifiers} \cite{sweeney2002k}. Therefore, with the aid of a small amount of external information, such as several credit card records \cite{de2015unique}, attackers are able to associate the anonymized trajectory with a single user, which realizes the $re$-$identification$ attack\cite{sweeney2002k}. More importantly, it is actually hard to alleviate mobile users' privacy leakage in anonymized mobility datasets. \cite{zang2011anonymization} and \cite{gramaglia2015hiding} tried to generalize or even permute the original mobility records before anonymization. Unfortunately, such operations require significant degeneration in data quality to achieve small benefit in preserving users' privacy.

Realizing the violation of user privacy in the releasing of anonymized mobility records, mobile operators now tend to release aggregated mobility datasets, such as the last three data release in Table~\ref{table:dataset}. For example, instead of sharing each anonymized record created by individuals, they now release the number of users covered by a base station, the number of voice calls made each hour in each location, etc. The operators believe that such aggregation will preserve users' privacy while providing useful statistic information for academic research and commercial usage.

\subsection{Privacy and Attack Model}

Our work targets for the privacy concerns in aggregated mobility data. The procedure to publishing aggregated mobility data can be summarized as following two steps: a) group the original mobility records of mobile users by time slots, b) compute and publish specific aggregated statistics of each time slot, e.g., the number of mobile users covered each base station. This privacy model is simple and effective since it ensures that the released datasets exhibit the following two desirable features:
\begin{itemize}
\item \textbf{No individual information can be directly acquired from the datasets.} Since the aggregated mobility data only contains general information of the population, we cannot directly distinguish each mobile user and extract personal information. It makes publishing aggregated mobility data automatically complies with $k$-$anonymity$ privacy model, which prevents the $re$-$identification$ attacks\cite{sweeney2002k}.
\item \textbf{The statistics of aggregated information is accurate.} The procedure of publishing aggregated mobility datasets does not require generalizing, suppressing or permuting original records. Therefore, it preserves the truthfulness and accuracy of the original datasets at record level. Such accurate aggregated mobility data is of great importance in numerous applications, ranging from transportation scheduling to business locating\cite{barabasi2005origin}.
\end{itemize}

Although the privacy leakage in publishing anonymized individual's mobility records has been recognized and extensively studied, the privacy issue in releasing aggregated mobility data remains unknown. To ensure the attack model's ability of generalization, we assume that the adversaries have no prior information of the targeted datasets, and define the attack model as recovering mobile users' individual trajectories from the aggregated mobility data with an unsupervised method. Once the adversaries accurately recover individual's trajectory, the privacy is under immediate threat of well-studied individual's records attack models, such as $re$-$identification$ attack and probability attack\cite{sweeney2002k, machanavajjhala2007diversity}, which have been proven to be effective in various scenarios\cite{de2013unique, zang2011anonymization}.

\section{Dataset and Fesibility}

\subsection{Mobility Dataset}

We use two real-world mobility datasets to understand and investigate how can we recover user's trajectory from the aggregated mobility data.

\mypara{Dataset collected by mobile application:}
This dataset is collected from mobile devices by a popular mobile application. It records the mobile users' spatiotemporal points when it is activated for service interactions. The dataset traces over 15,000 mobile users from November 1st to 14th, 2015. It records fine-grained spatiotemporal information of mobile users, including anonymized user identification, accessed base stations and timestamp.

\mypara{Dataset collected by cellular operator:} This dataset is collected by a major mobile network operator in China. It is a large-scale dataset including 100,000 mobile users with the duration of one week, between April 1st and 7th, 2016. It records the spatiotemporal information of mobile subscribers when they access cellular network (i.e., making phone calls, sending texts, or consuming data plan). It also contains anonymous user identification, accessed base stations and timestamp of each access.

\begin{table}[t]
\begin{center}
\begin{tabular}{c|cc}
\toprule
{\scriptsize Datasets \& Metrics}  &  \scriptsize  Operator Dataset & \scriptsize  Application Dataset  \\
\midrule
\scriptsize Source    &\scriptsize Cellular network &  \scriptsize Mobile application   \\
\scriptsize Time & \scriptsize  Apr. 2016& \scriptsize Nov. 2015 \\
\scriptsize Duration & \scriptsize one week & \scriptsize two weeks\\
\scriptsize  User number   & \scriptsize 100,000 & \scriptsize 15,500  \\
\scriptsize Average records per user & \scriptsize 261 & \scriptsize 496 \\
\bottomrule
\end{tabular}
\end{center}
\caption{Key features of our utilized two different mobility datasets.}\label{tab:data}
\end{table}
\normalsize

Both datasets are collected in a major city with over 8,000 base stations in China.
Therefore, by looking up the locations of these base stations, we are able to obtain trajectories of mobile users, which serve as the ground truth in the investigation of aggregated mobility data's privacy leakage. Table~\ref{tab:data} summarized the key features of these two datasets. The number of users and average number of records per user are diverse in these two datasets, which makes our investigation covering a broad range of scenarios in dataset releasing. We want to point out that we have taken the following steps to ensure the ethical considerations of dealing with such sensitive data: first, all mobile users' identifiers are replaced with random sequences to achieve anonymizations; second, we store all the data in a secure local server; third, only the core researchers regulated by the strict non disclosure agreements have access to the data.



\subsection{Why Privacy is NOT Preserved?}\label{sec:feasibility}

\begin{figure}[t]
\begin{center}
\subfigure[Day 1]{\includegraphics[width=0.235\textwidth]{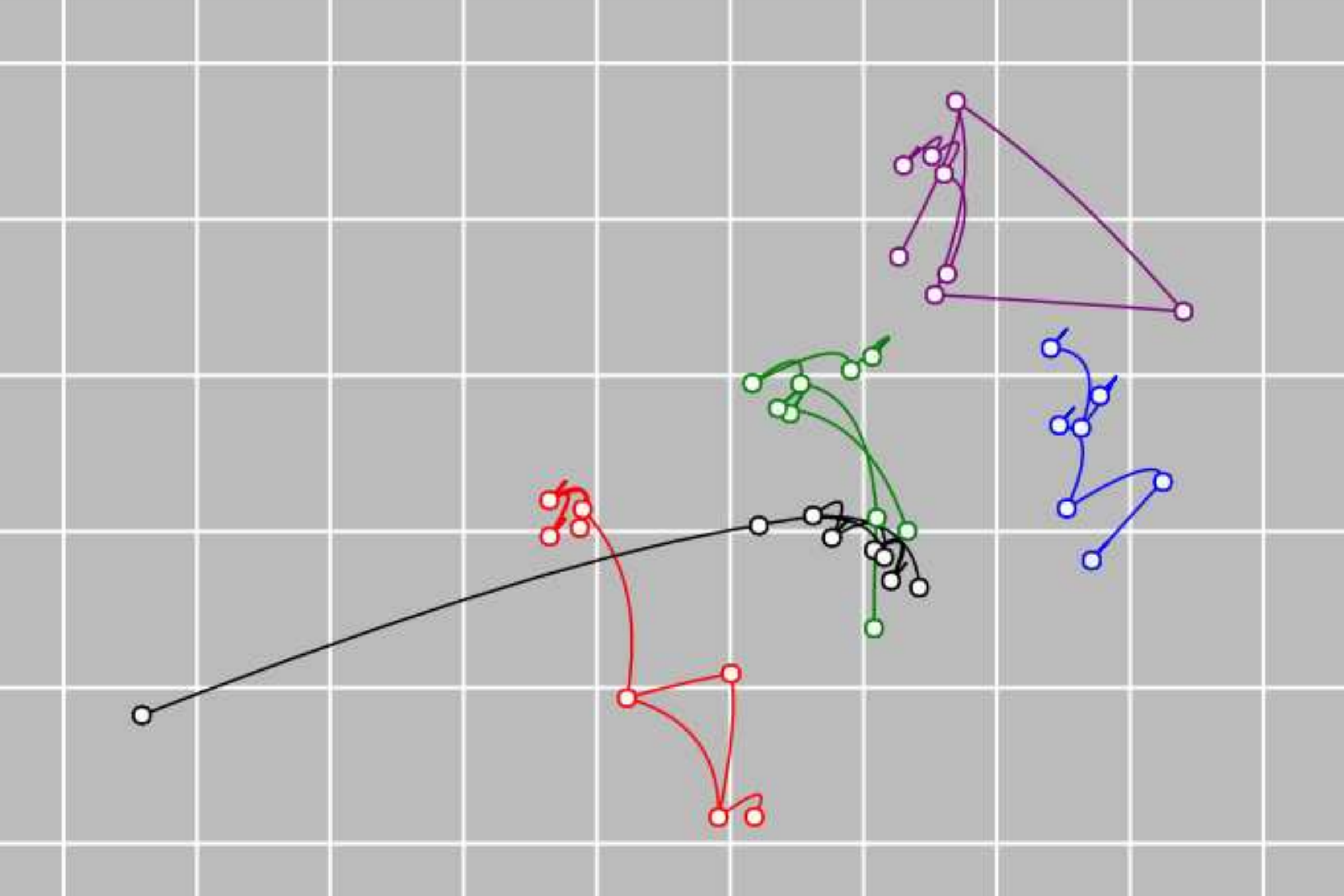}}
\subfigure[Day 2]{\includegraphics[width=0.235\textwidth]{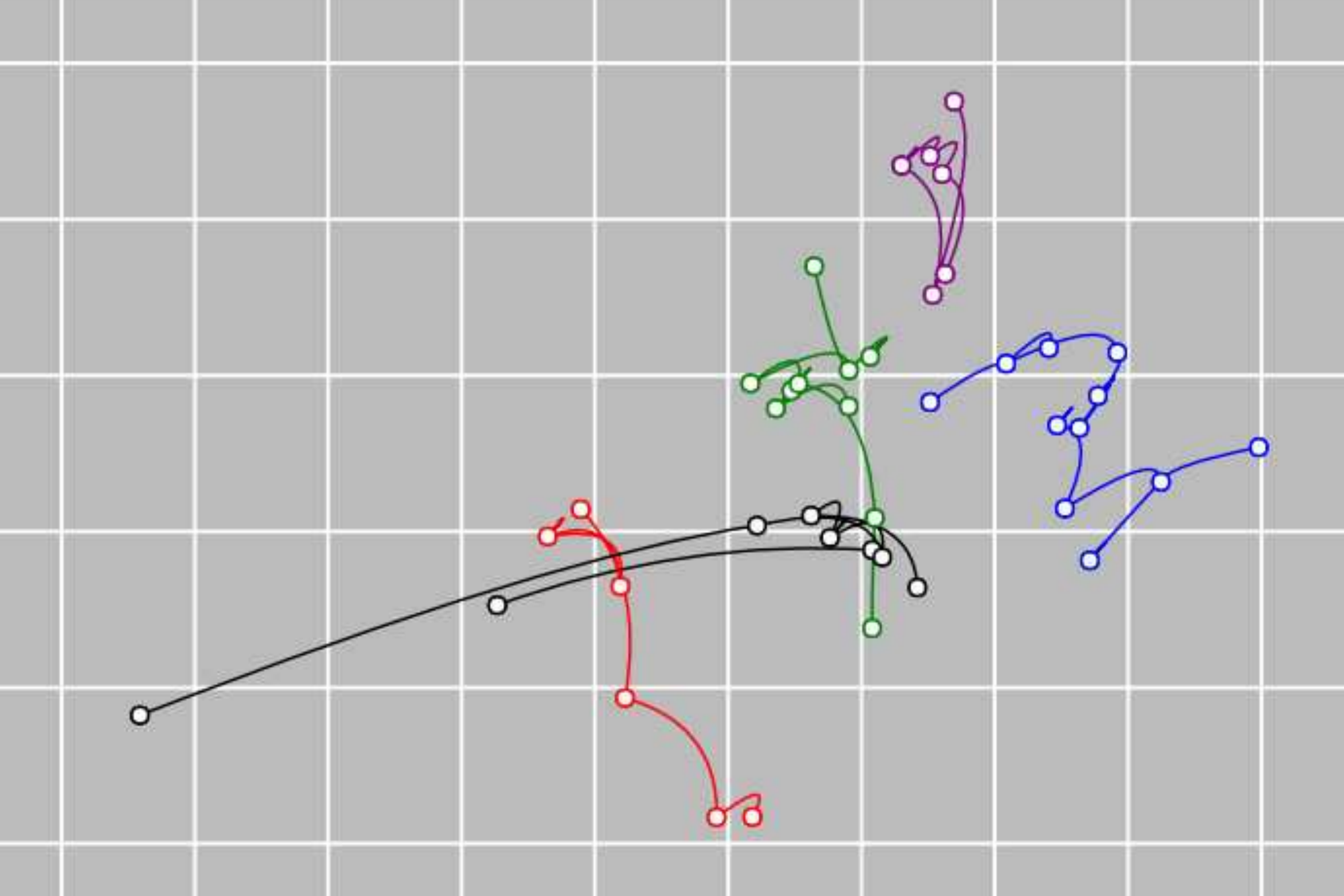}}
\caption{Mobility trajectories of five randomly selected mobile users in two days.}
\label{fig:trajectories}
\end{center}
\end{figure}

We argue that aggregating mobility records does not preserve users' privacy, since a user's mobility pattern is regular while different from others'. Figure~\ref{fig:trajectories} shows the mobility trajectories of five randomly selected users from the operator dataset over two days. We can clearly observe that each user has a coherent mobility trajectory, i.e., their mobility trajectories in the first day are similar to that of the second day. More importantly, they are significantly different from each other. As a result, even though the data owners like cellular operators release aggregated mobility datasets, a malicious entity still has the opportunity of recovering each user's trajectory by exploiting such regularity and uniqueness. One natural question to ask is that does every mobile user have such coherent and unique mobility trace?

\begin{figure}
\centering
\subfigure[Operator dataset]{\includegraphics[width=0.235\textwidth]{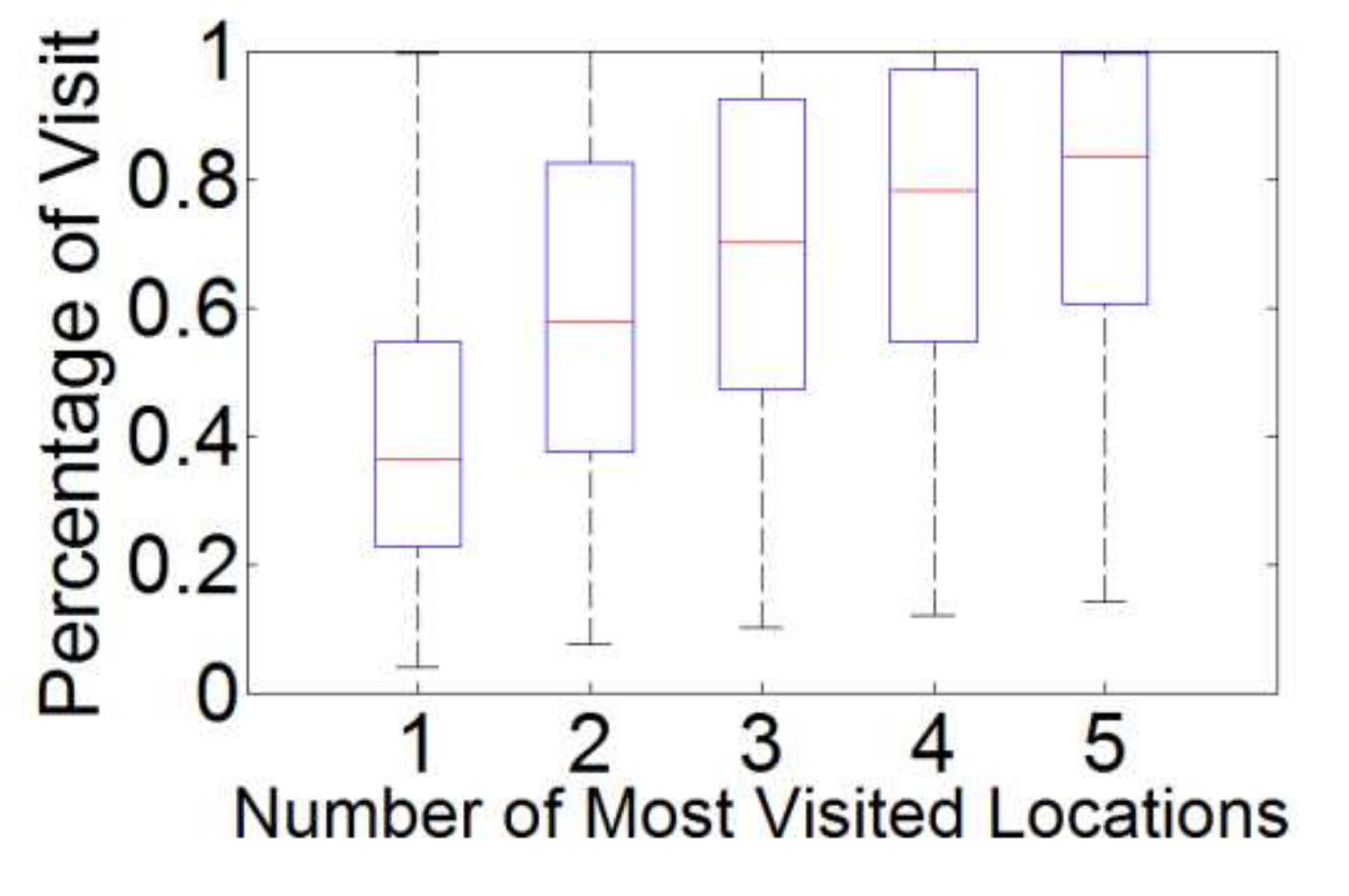}}
\subfigure[App dataset]{\includegraphics[width=0.235\textwidth]{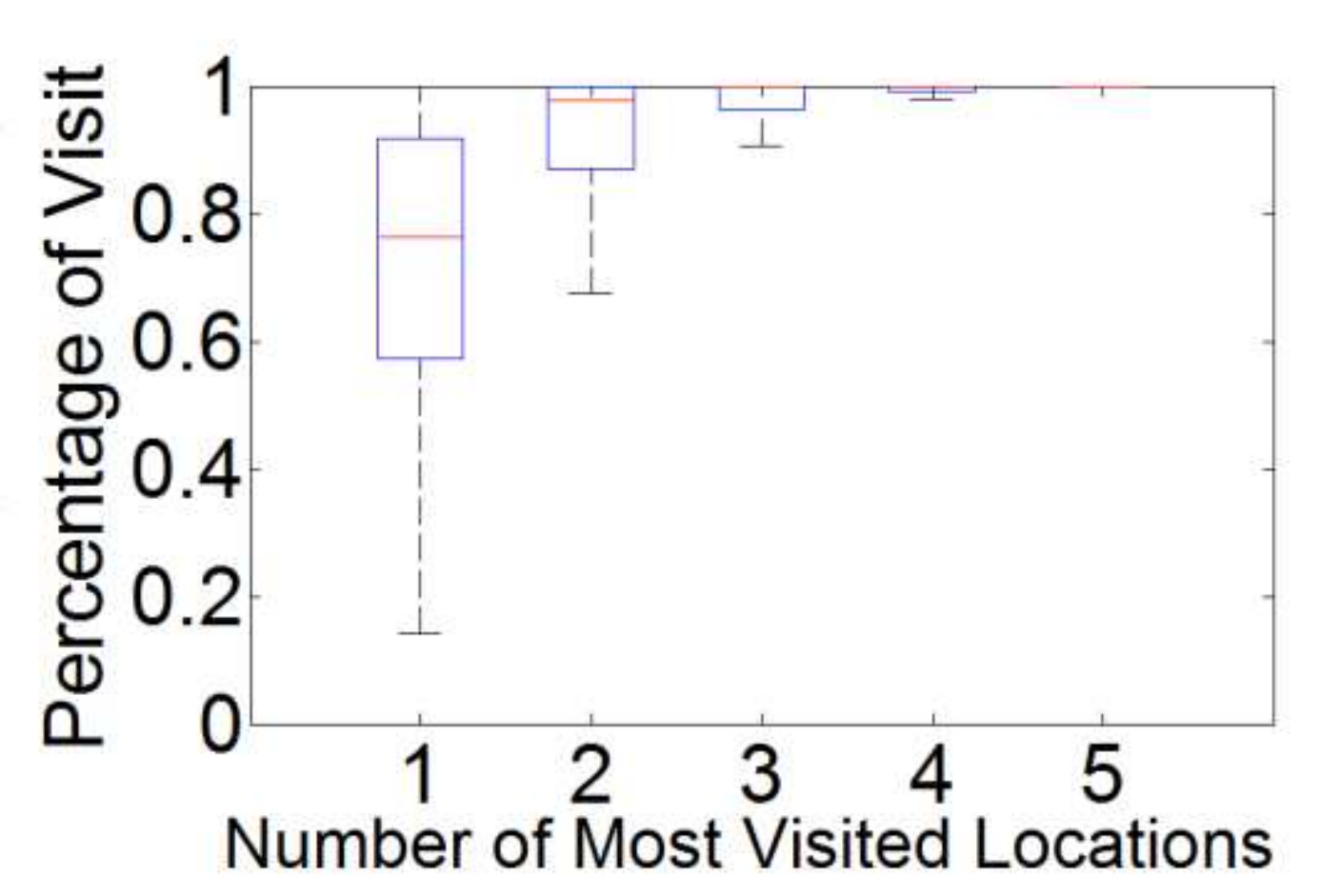}}
\caption{The percentage of mobility records happen in each mobile user's most frequently visited towers.}
\label{fig:heat_pop_user}
\end{figure}

To answer this question, we evaluate the regularity and uniqueness of all mobile users' trajectories on both investigated datasets. First, we present the percentage of mobility records that happen in each mobile user's most frequently visited cellular towers in Figure~\ref{fig:heat_pop_user}. For the operator dataset, Figure~\ref{fig:heat_pop_user}(a) shows that 36\% of records happen in the most frequent (top one) cellular tower during that week. Similarly, on average a user visits top five towers for 83\% of the time. Both of them suggest that a user always stays in the top towers for most of his/her daily cellular access. We have similar observation on the other dataset shown in Figure~\ref{fig:heat_pop_user}(b). Mobile users visit their top one towers for 76\% of the time and almost always stay in their top five towers. Such mobility pattern comes from the fact that mobile users usually stay at one or several locations during their daily day, such as office and home. As a result, only limited number of towers are visited. These results reveal that mobile users' mobility patterns are highly regular, which is consistent with the coherent mobility trajectories observed in Figure~\ref{fig:trajectories}.

\begin{figure}
\centering
\subfigure[Operator dataset]{\includegraphics[width=0.235\textwidth]{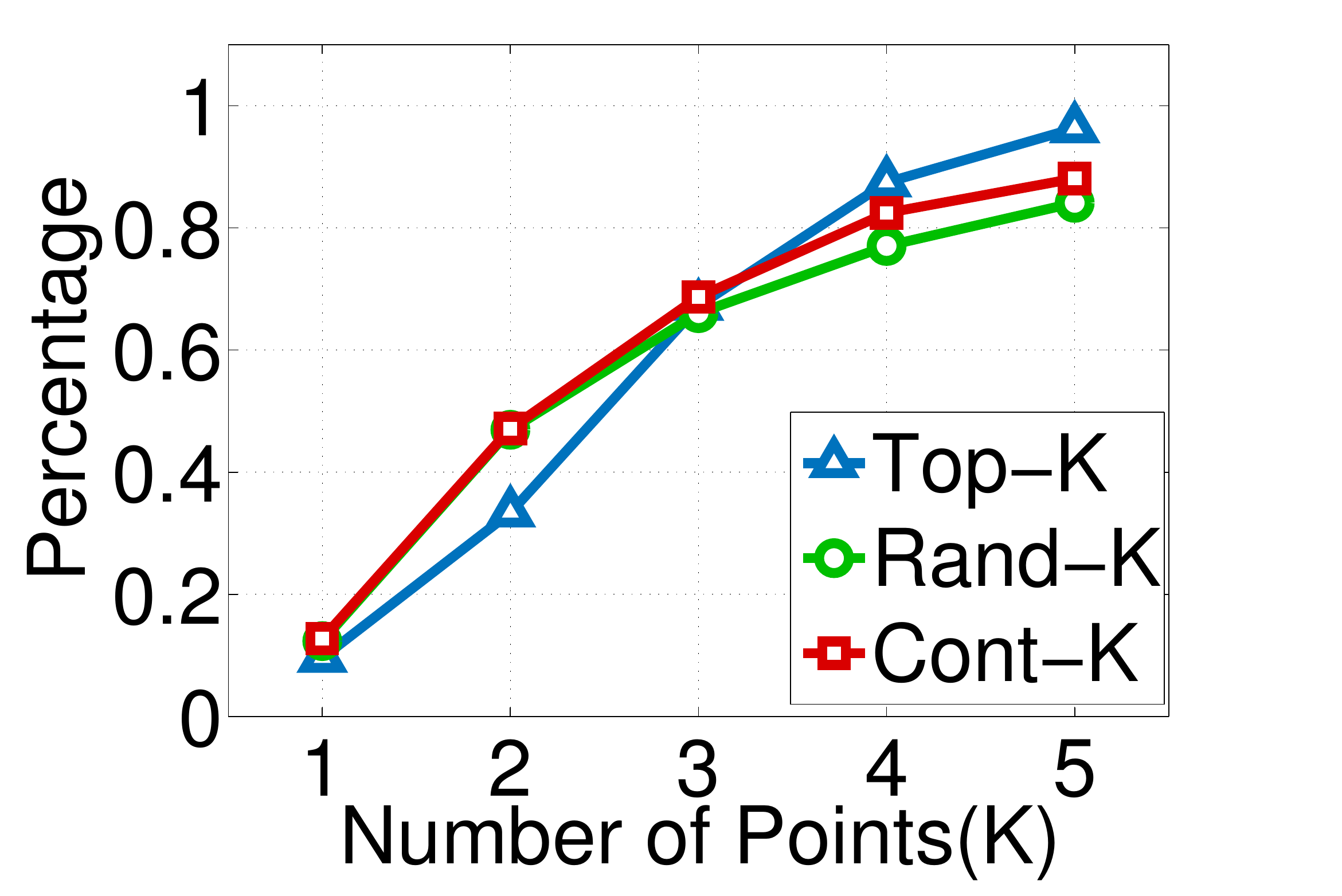}}
\subfigure[App dataset]{\includegraphics[width=0.235\textwidth]{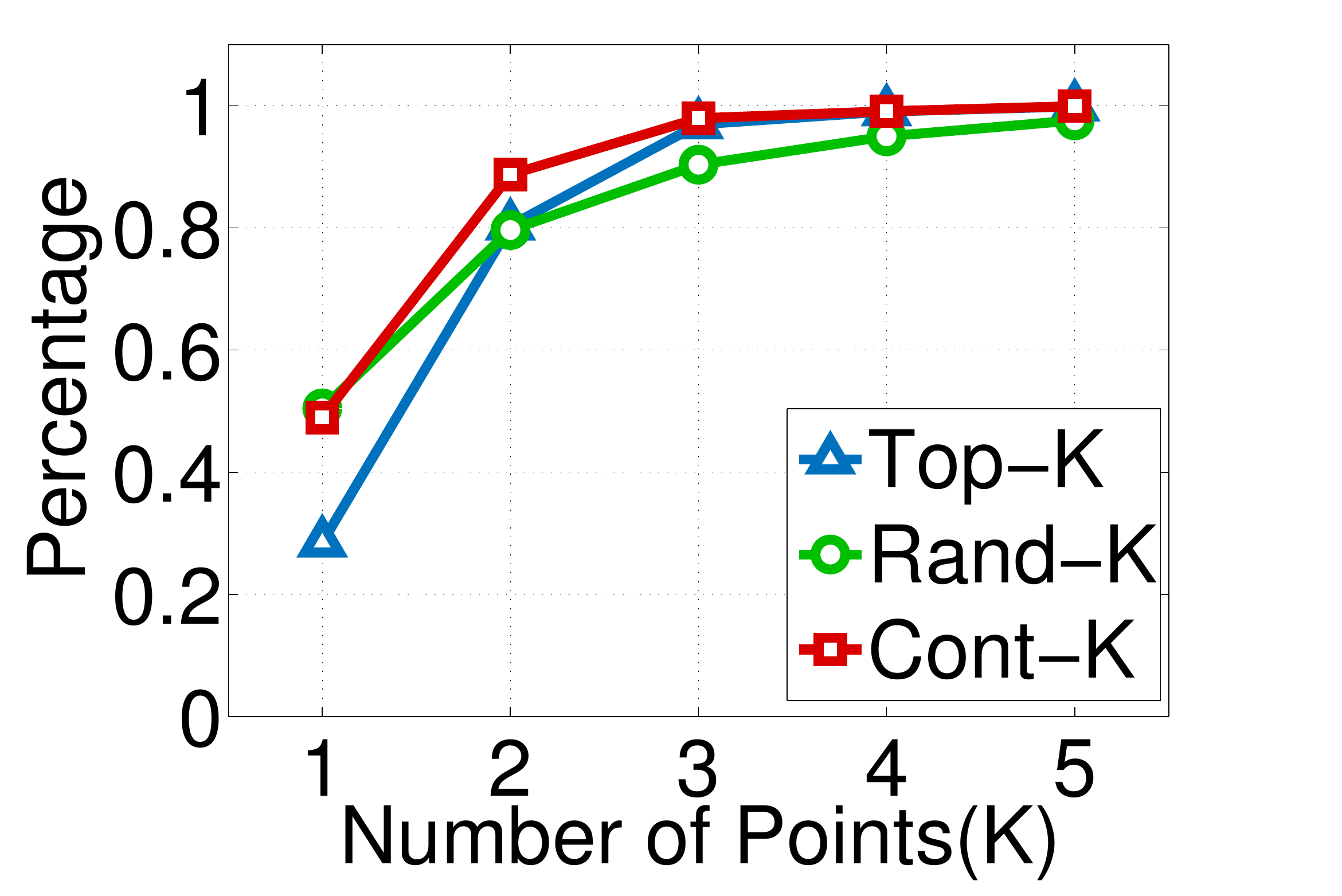}}
\caption{The percentage of mobile users that can be distinguished by the selected $K$ spatiotemporal points. }
\label{fig:N points}
\end{figure}

In order to quantify the uniqueness of mobile user's trajectory, we investigate the possibility of distinguishing different users by choosing only a small set of the spatiotemporal records in their trajectories. We use three different strategies to obtain the sets of spatiotemporal points, i.e., selecting the top $K$ frequented locations visited by the user (Top-$K$), randomly selecting $K$ spatiotemporal points belong to the user's trajectory (Rand-$K$), and randomly selecting $K$ consecutive spatiotemporal points (Cont-$K$). Under these three strategies, we show the percentage of mobile users that can be distinguished from others by the selected records in Figure~\ref{fig:N points}. In the operator dataset, we can observe that 67\% users can be distinguished when we look at the Top-$3$ towers. When we look at the Top-$5$ towers, above 95\% of users are unique. As for the other two strategies, we can observe higher percentage of unique users, when the number of points is less than 3. Similar observation can be found in the application dataset. All these results quantitatively demonstrate that the individual mobility is significantly different across different users, and most of them are unique. Based on the uniqueness and the regularity of users' mobility, we come to the conclusion that it is possible to recover mobile user's trajectory from the aggregated mobility dataset.

\section{Mobility Trajectory Recovery}

We design an unsupervised framework that leverages the universal characteristics of human mobility to recover users' trajectories from aggregated mobility data without any prior knowledge. Our framework includes three modules: nighttime, daytime, and cross-day trajectory recovery. After properly formulating the investigated problem, we discuss each module with details.

\subsection{Problem Definition}

In our considered privacy model, we release aggregated spatiotemporal information of mobile users, i.e., the number of mobile users covered by each cellular tower at each time slot. Formally, we define such data as $P^t = [p_1^t, p_2^t,..., p_M^t]$, with $p_m^t$ representing the number of mobile users at location $m$ in time slot $t$ and $M$ representing the total number of locations. Based on the number of each location's mobile users, we can directly derive the ID removed mobility records $L^t = [l_1^t, l_2^t,..., l_N^t]$ at time slot $t$ with $N$ as the total number of mobile users. Recovering a user's mobility trajectory is equivalent to associating the ID removed mobility records that are created by the same user across different time slots. Therefore, the fundamental question to ask is: how to identify the mobility records that belong to the same mobile users?

To answer this problem, we propose an attack framework that iteratively associates the same users' mobility records in the neighbouring time slots, and step by step recovers the whole trajectories. In each time slots, the procedure of attack system can be broke down into two steps. First, estimate the likelihood of next location $l_i^{t+1}$ belongs to a given trajectory by exploiting the characteristics of human mobility. Second, derive an optimal solution to link mobile users' trajectories with next mobility records that maximize the overall likelihood. We first focus on the second step --- given estimated likelihood, how to derive an optimal association.

Formally, we define the recovered trajectories till time slot $t$ as $S^t = [s^t_1, s^t_2,..., s^t_N]$, where $s^t_j = [q^1_j, q^2_j,..., q^t_j]$ is the $j$th recovered trajectory and $q^t_j$ is the recovered location at time slot $t$. Given the cost matrix $C^t = \{c^t_{i,j}\}_{N\times N}$, with $c^t_{i, j}$ corresponding to the inverse of likelihood of linking trajectory $s^t_i$ with next location $l^{t+1}_j$, the trajectory recovery problem is equivalent to finding the optimal match between the rows and columns with least overall cost. Formally, we define the decision matrix $X^t = \{ x^t_{i, j} \}_{N\times N}$, with $x^t_{i, j} = 1$ denoting linking trajectory $s^t_i$ with location $l^{t+1}_j$ and $x^t_{i, j} = 0$ otherwise. Then, the trajectory recovery problem can be formulated as following,

\begin{equation}
\begin{aligned}
Minimize &\sum_{i=1}^N \sum_{j=1}^N c^t_{i, j} \times x^t_{i, j}, \\
subject\ to&\ x^t_{i, j} = \{0, 1\}, \ \sum_{i=1}^N x^t_{i, j}=1, \ \sum_{j=1}^N x^t_{i, j}=1.
\end{aligned}
\end{equation}

The above formulated problem is equivalent to \emph{Linear Sum Assignment Problem}\cite{dantzig1998linear}, which has been extensively studied and can be solved in polynomial time with \emph{Hungarian algorithm}\cite{kuhn1955hungarian}. The main procedure of this algorithm is following,

\begin{itemize}
\item Step 1: In cost matrix $C^t$, the elements of each row are subtracted with the smallest element in that row. Then, the elements of each column are subtracted with the smallest element in that column.
\item Step 2: Draw minimum number of lines through rows and columns to cover all the zero entries in $C^t$.
\item Step 3: a)If the number of covering lines is $N$, derive the optimal match among the zero entries and stop the algorithm. b) If the number of covering lines is less than $N$, proceed to Step 4.
\item Step 4: Find the minimum entry that is not covered by the lines. Subtract each element of uncovered rows with this entry, and then add each element of the covered column with this entry.  Return to Step 2.
\end{itemize}

Therefore, the remaining problem of trajectory recovery is how to build accurate cost matrix $C^t$ by exploiting mobility characteristic. Explicitly, we design three different schemes to estimate the cost matrix $C^t$ at different time periods: nighttime, daytime and across days, which are elaborately discussed as follows.

\subsection{Recovering Nighttime Trajectories }

\begin{figure}
\centering
\subfigure[Operator dataset]{\includegraphics[width=0.235\textwidth]{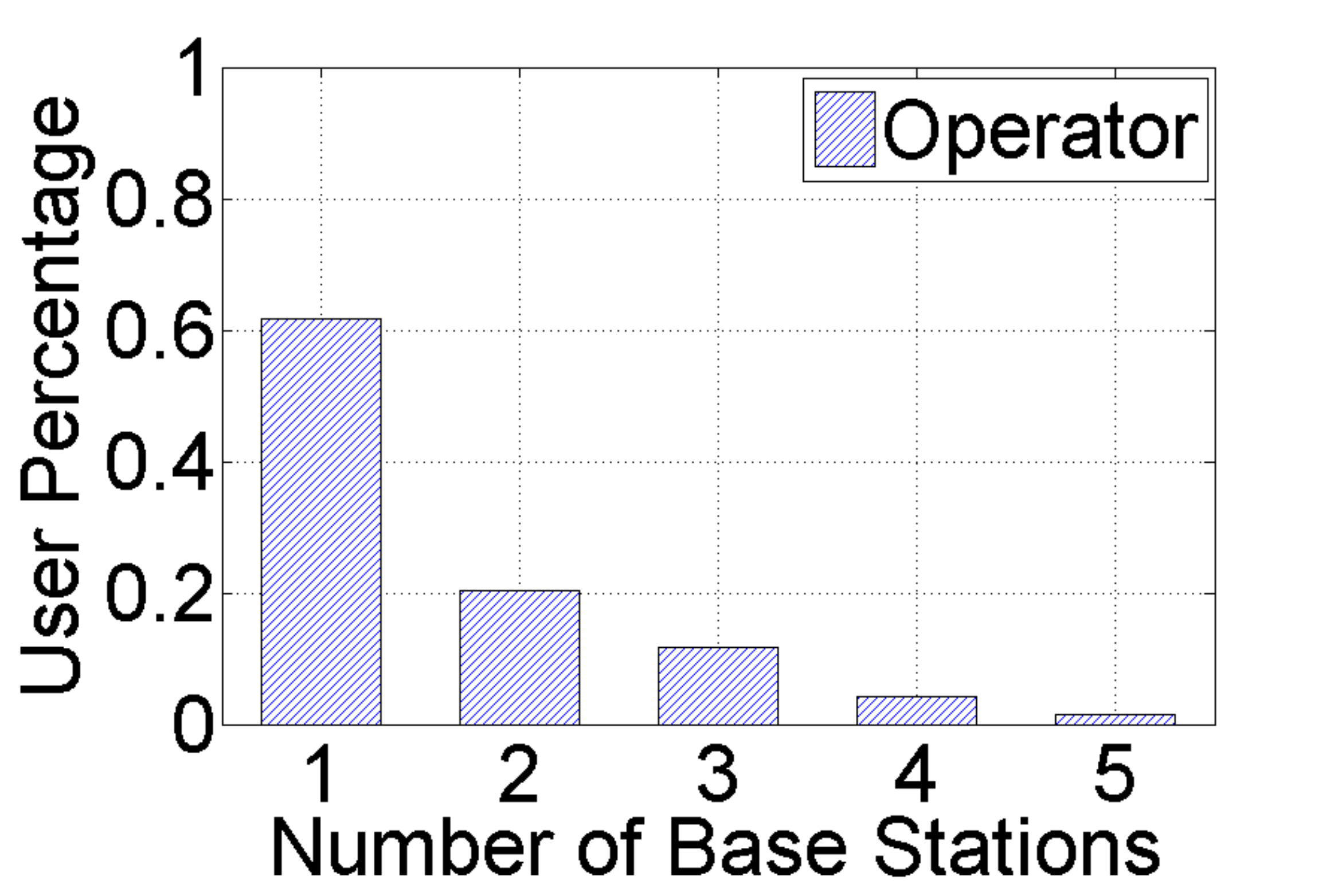}}
\subfigure[App dataset]{\includegraphics[width=0.235\textwidth]{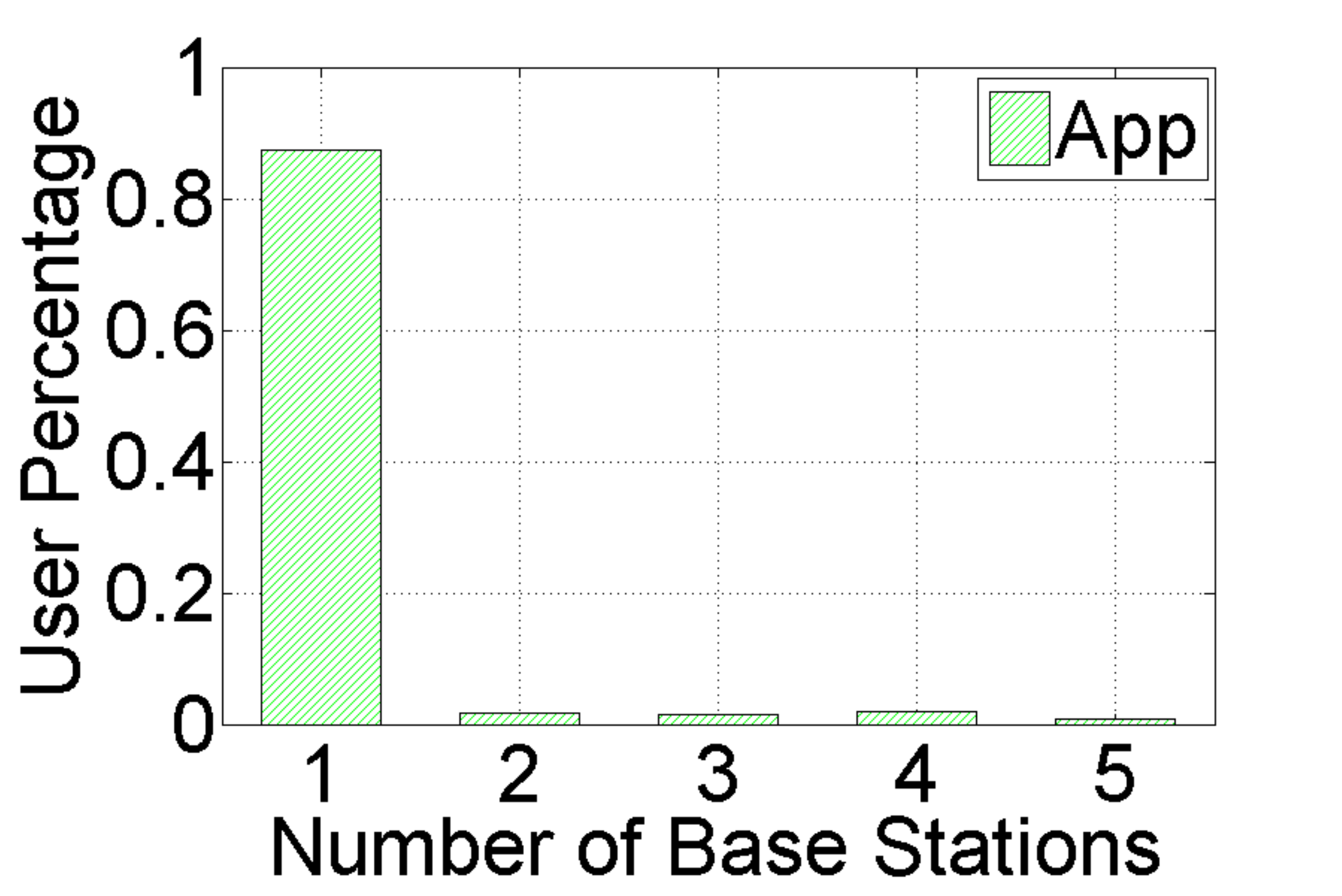}}
\caption{Percentage of users with different number of nighttime locations.}
\label{fig:time_interval_distribution}
\end{figure}

\begin{figure}
\centering
\subfigure[Operator dataset]{\includegraphics[width=0.235\textwidth]{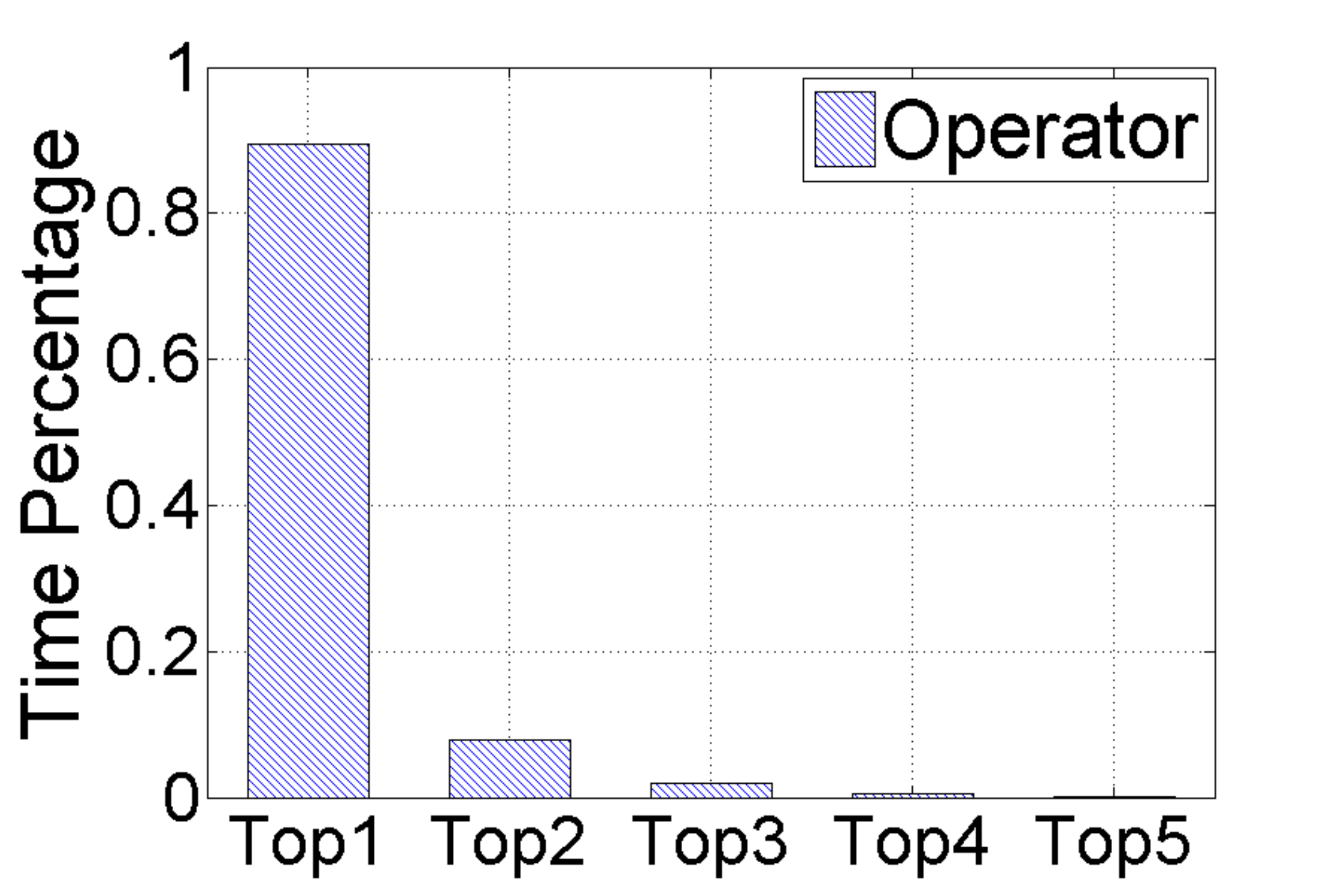}}
\subfigure[App dataset]{\includegraphics[width=0.235\textwidth]{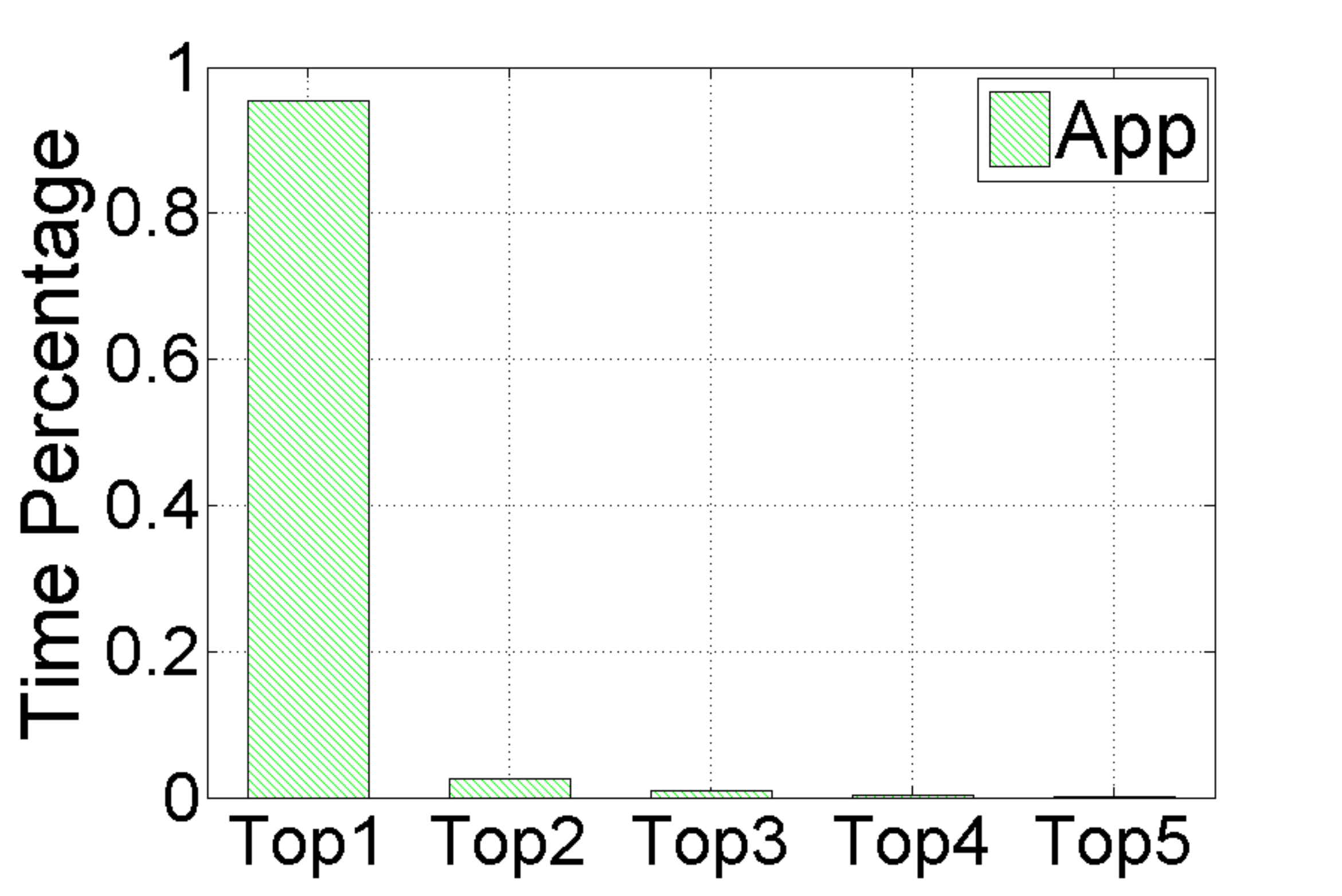}}
\caption{Percentage of time staying in the most frequent locations during nighttime.} \label{fig:distance_distribution}
\end{figure}

The key insight of nighttime trajectory recovery is that mobile users tend to stay in fixed locations during nighttime for natural sleeping cycle. Figure~\ref{fig:time_interval_distribution} shows the percentage of users with different number of visited locations during nighttime, i.e., 0 am$\sim$6 am. From the results, we can observe that 62\% and 88\% of mobile users only visit one base station during nighttime in operator and application datasets, respectively. On the other hand, Figure~\ref{fig:distance_distribution} shows the percentage of nighttime that mobile users stay in their top frequent base stations. In the operator dataset, 89\% of nighttime is spent in the most frequent base station, while the percentage reaches to 95\% in the application dataset. These results demonstrate that mobile users are of low mobility during nighttime and tend to stay in the same base stations.

Based on above observations, we design a scheme to generate the cost matrix $C^t$ during nighttime by following two steps: a) estimate the next location of each recovered trajectory as the last location visited by that trajectory. Defining the estimated next location at time slot $t+1$ of the $i$th recovered trajectory as $\hat l^{t+1}_i$, the estimated next locations is derived by $\hat l^{t+1}_i = q^t_i$; b) use the distance between the estimated next location $\hat l^{t+1}_i$ and the actual location point $l^{t+1}_j$ as the cost $c^t_{i, j}$ of linking $l^{t+1}_j$ to trajectory $s^t_i$. In this way, we define the cost matrix $C^t$ aiming to minimize the estimation error.

Based on the above defined $C^t$, we then utilize \emph{Hungarian algorithm} to achieve optimal association between recovered trajectories and next location points, which recovers the nighttime trajectories.

\subsection{Recovering Daytime Trajectories}

\begin{figure}[t]
\centering
\subfigure[Operator dataset]{\includegraphics[width=0.235\textwidth]{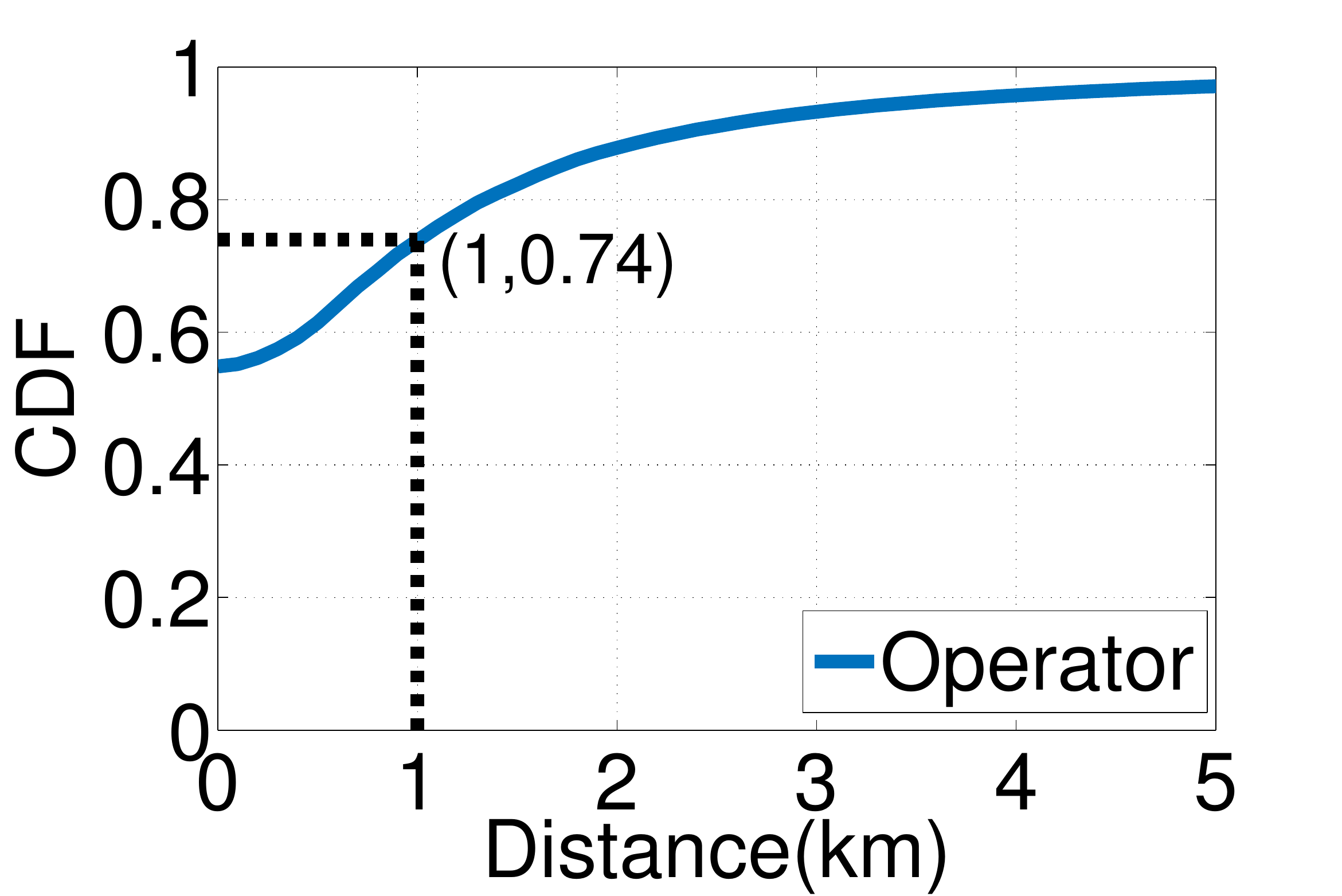}}
\subfigure[App dataset]{\includegraphics[width=0.235\textwidth]{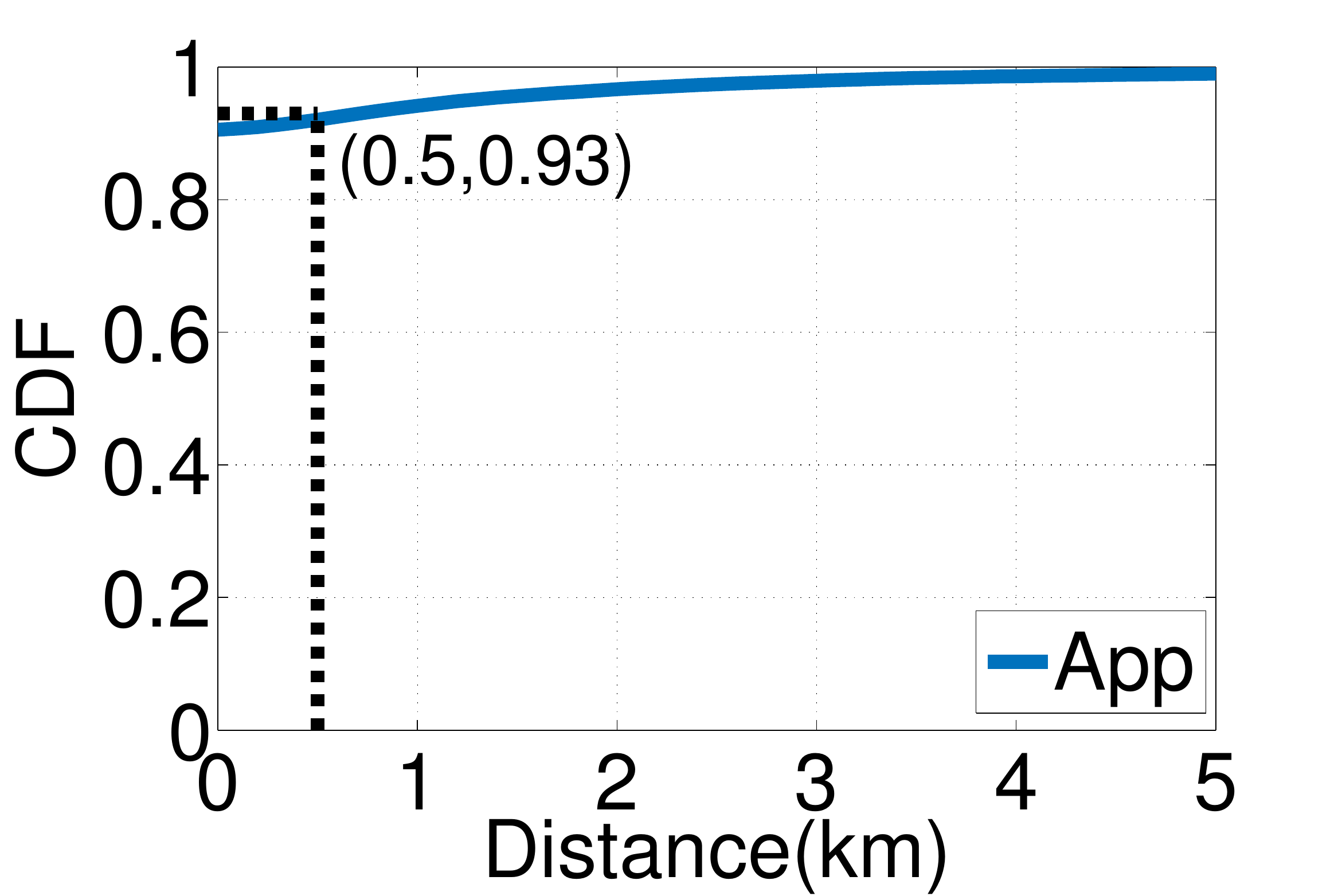}}
\caption{The CDF of the errors between the predicted location and ground truth.} \label{fig:gauss}
\end{figure}

Different from nighttime, users move frequently during daytime, which requires a different scheme to estimate the next locations. The key insight is the continuity of human mobility, which enables the next location estimation by using the current location and velocity. Our estimation model is as follows,
\begin{equation}
\label{eq:velocity}
\hat l_i^{t+1} =  q^t_i + (q^t_i - q^{t-1}_i),
\end{equation}
where $q^t_i$ is the location at timestamp $t$ of the $i$th trajectory. Since the locations of mobile users are reported periodically, $q^t_i - q^{t-1}_i$ is the displacement between timestamp $t$ and $t+1$ estimated by the velocity of last movement.

Figure~\ref{fig:gauss} shows the empirical cumulative distribution function (CDF) of the errors between the prediction and ground truth location. From the results, we can observe that the estimated locations are accurate in terms of 74\% of them have an error less than 1,000 meters in operator dataset, while 93\% of them have less than 500 meters error in application dataset. Therefore, we use the distance between the estimated next locations and the unassigned mobility records in next time slot to formulate the cost matrix $C^t$, where $c^t_{i, j}= distance(\hat l_i^{t+1}, l_j^{t+1})$.

By applying \emph{Hungarian algorithm} on matrix $C^t$, we link the trajectories with next locations by maximizing the likelihood. Since minimizing distance is equivalent to maximizing the probability, we can iteratively link the recovered nighttime trajectories with aggregated records, and recover the trajectories of each day.

It worth noting that \emph{Hungarian algorithm} is currently the most efficient algorithm to solve this problem, but still with computational complexity of $O(n^3)$. To speed up, we adopt a suboptimal solution to reduce the dimension of cost matrix by taking out the pairs of trajectories and location points with cost below a predefined threshold and directly link them together. This approach effectively reduces the running time of trajectory recovery, while keeps a good performance at the same time.

\subsection{Recovering Trajectories Across Days}
\begin{figure}[t]
\centering
\subfigure[Operator dataset]{\includegraphics[width=0.235\textwidth]{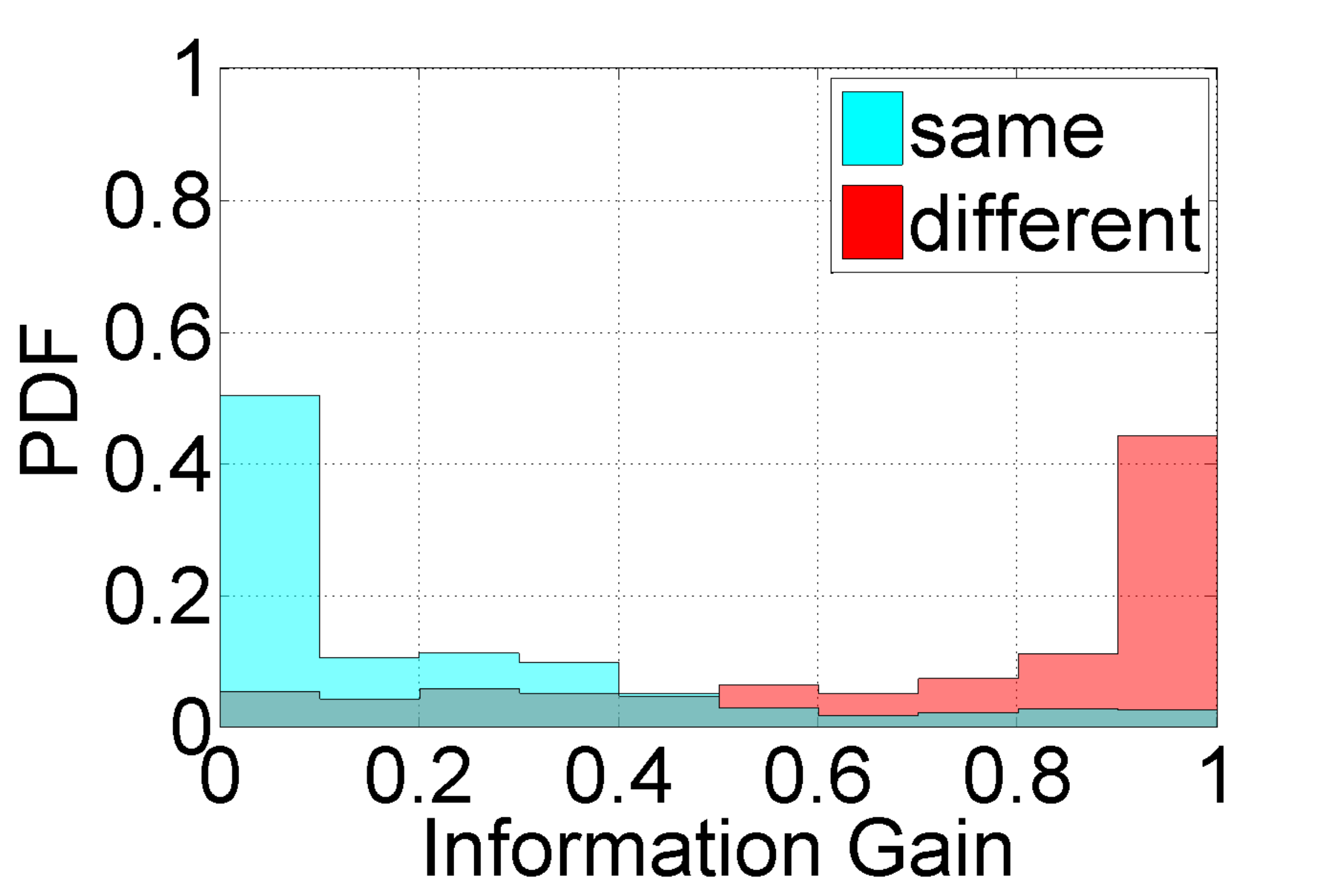}}
\subfigure[App dataset]{\includegraphics[width=0.235\textwidth]{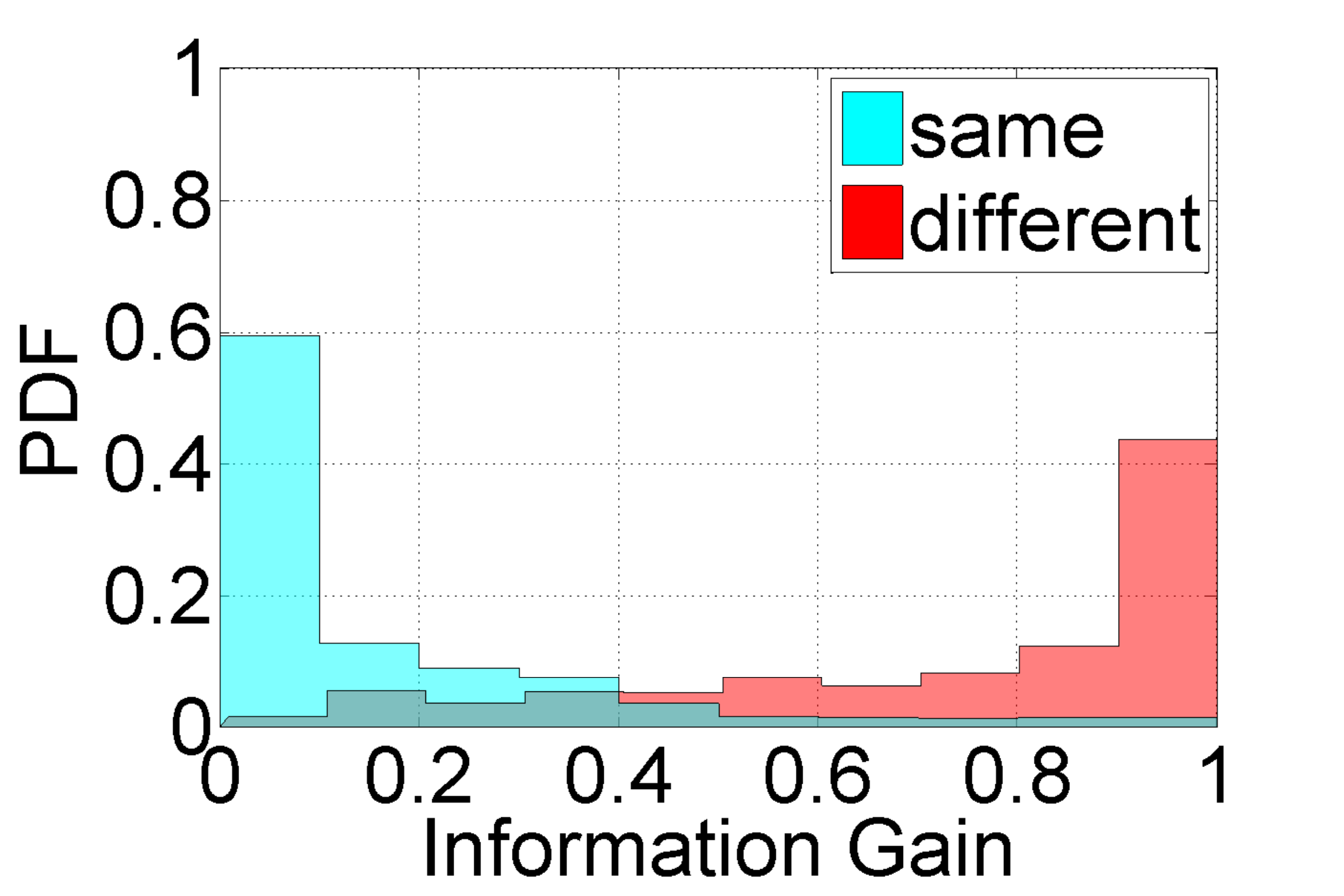}}
\caption{The PDF of information gain in linking across days sub-trajectories contributed by a single user or different users.} \label{fig:information}
\end{figure}

After above two steps, we have recovered mobile users' sub-trajectories of each day. To associate the sub-trajectories that belong to same users, we leverage two important features that a user's mobility pattern is regular and different users have significantly different patterns. Specifically, we use the \emph{information gain} of connecting two sub-trajectories to measure their similarities. We define the $i$th recovered sub-trajectories in the $d$th day as $U_i^d$. Since the information gain is designed based on entropy, we first introduce the formulation of the entropy as follows,

\begin{equation}
\label{eq:entropy}
H(U_i^d) = -\sum_{k=1}^n{\frac{f_k}{\sum_{k=1}^n{f_k}}\log(\frac{f_k}{\sum_{k=1}^n{f_k}})},
\end{equation}
where $f_k$ is the frequency of visiting the $k$th cellular tower in sub-trajectory $U_i^d$, and $H(U_i^d)$ is entropy of that sub-trajectory. We denote the information gain of linking two sub-trajectories $U_i^d$ and $U_j^{d+1}$ as $G(U_i^d,\ U_j^{d+1})$, which is computed as follows,
\begin{equation}
\label{eq:info_gain}
G(U_i^d,\ U_j^{d+1})= H(U_i^d + U_j^{d+1})- \frac{H(U_i^d) + H(U_j^{d+1})}{2},
\end{equation}
where $H(U_i^d + U_j^{d+1})$ is the entropy of the combined trajectory. The information gain measures the difference of frequency distribution over towers recorded in $U_i^d$ and $U_j^{d+1}$. If $U_i^d$ and $U_j^{d+1}$ have similar  distribution on cellular towers, the information gain will be close to zero. Otherwise, it will be close to one. Figure~\ref{fig:information} shows the probability distribution function (PDF) of the information gain when we associate two sub-trajectories that contributed by the same users as well as the PDF of associating different users' sub-trajectories. We can observe that these two PDFs are significantly different in both datasets. We obtain close to zero information gain when we group two sub-trajectories contributed by the same user. In contrast, close to one information gain is obtained when we combine two sub-trajectories of different users. Therefore, we use the information gain of combining two sub-trajectories to generate the cost matrix, where $c^d_{i, j}=G(U_i^d,\ U_j^{d+1})$. By solving the optimal match problem, we associate the sub-trajectories with highest similarity across different days, and recover the whole trajectories of each mobile user.

To conclude, we design an unsupervised attack framework that utilizes the universal characteristics of human mobility to recover individual's trajectory in aggregated mobility datasets. Since the proposed framework does not require any prior information of the target datasets, it can be easily applied on other aggregated mobility datasets. Note that we do not aim to design an optimal while sophisticate attack system. Instead, we intend to build a elementary but effective attack system to reveal the privacy leakage in aggregated mobility datasets.


\section{Performance Evaluation} \label{sec:evaluation}

In this section, utilizing the two introduced datasets, we investigate the possibility of privacy leakage from the aggregated mobility data by applying our designed attack system. We first discuss how we extract ground truth from the datasets, then define metrics to measure the privacy leakage, and finally conduct extensive evaluations to quantify them.

\subsection{Data Preparation and Metrics}
\textbf{Ground Truth.} Our two real-world datasets contain spatiotemporal records of each individual's mobility trajectory. Since we target aggregated mobility data, we perform the standard aggregating procedures on the datasets to obtain the ground truth\cite{blondel2012data,acs2014case,dengta}. Such procedure includes three steps: \#1) Group all  records in each 30 minutes time slot. \#2) Extract the most frequently visited location of each mobile user in each time slot, and perform a linear interpolation to determine the locations without records. \#3) Aggregate the extracted locations and generate the number of mobile users covered by each cellular tower in each time slot.

\textbf{Performance Metrics.} We introduce three metrics to evaluate the performance of our attack system ---\emph{accuracy}, \emph{recovery error} and \emph{uniqueness}. To properly measure these metrics, we first uniquely pair the recovered trajectories with the most similar trajectories in the ground truth data. The pairing is done in a greedy manner by enumerating the recovered trajectories and finding the most similar original trajectories that are not paired. Intuitively, the privacy leakage is severe, if the recovered trajectories are of high accuracy, i.e., most of the recovered spatiotemporal points are correct. Therefore, we compute the accuracy as the ratio between the number of correctly recovered spatiotemporal points and the total number. Denoting the $i$th original trajectory as $Y_i = [y^1_i, y^2_i,..., y^T_i]$ and the $i$th recovered trajectory that is paired with $Y_i$ as $Z_i = [z^1_i, z^2_i,..., z^T_i]$ with $T$ as the total number of time slot, we compute the recover accuracy, denoted by $A$, as follows,
\begin{equation}
A = \frac{1}{N}\sum_{i=1}^N {\frac{| Z_i \cap Y_i |}{|Y_i|}},\\
\end{equation}\\
where $Z_i \cap Y_i$ is the common spatiotemporal points between the two trajectories and $|\star|$ is the number of spatiotemporal points in $\star$.

\begin{figure*}[t]
\centering
\subfigure[Recovery accuracy]{\includegraphics[width=0.325\textwidth]{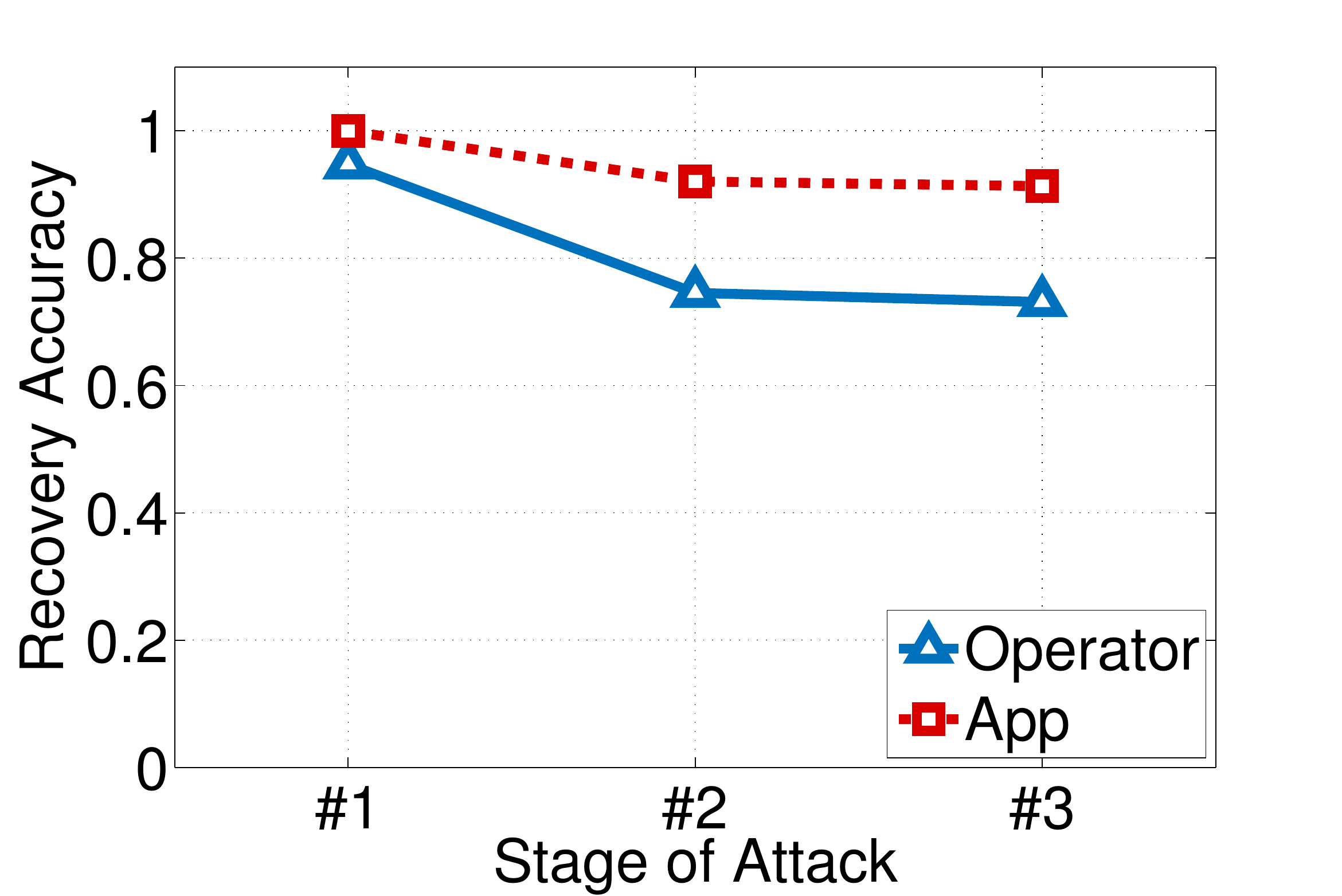}}
\subfigure[CDF of recovery error]{\includegraphics[width=0.325\textwidth]{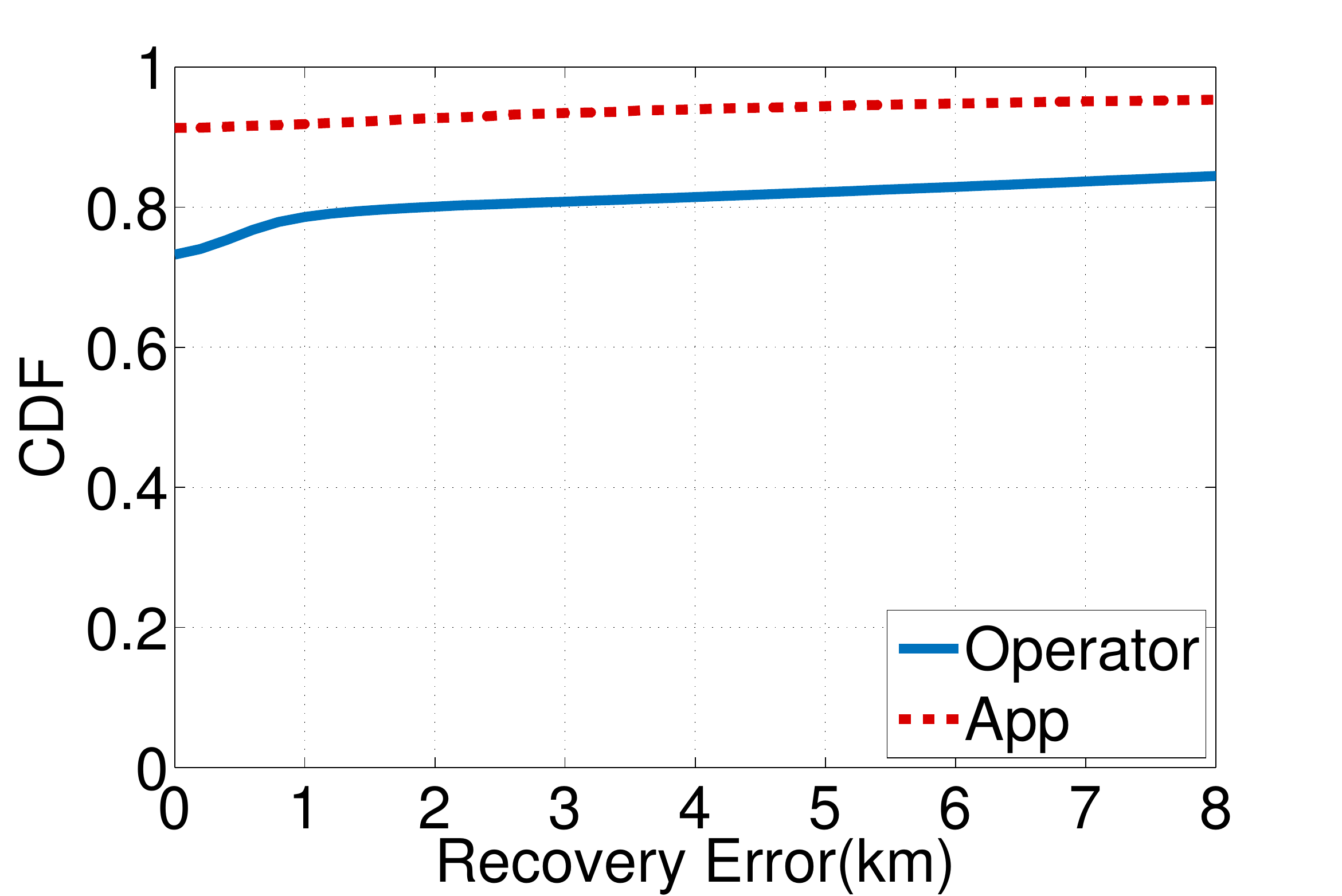}}
\subfigure[Uniqueness]{\includegraphics[width=0.325\textwidth]{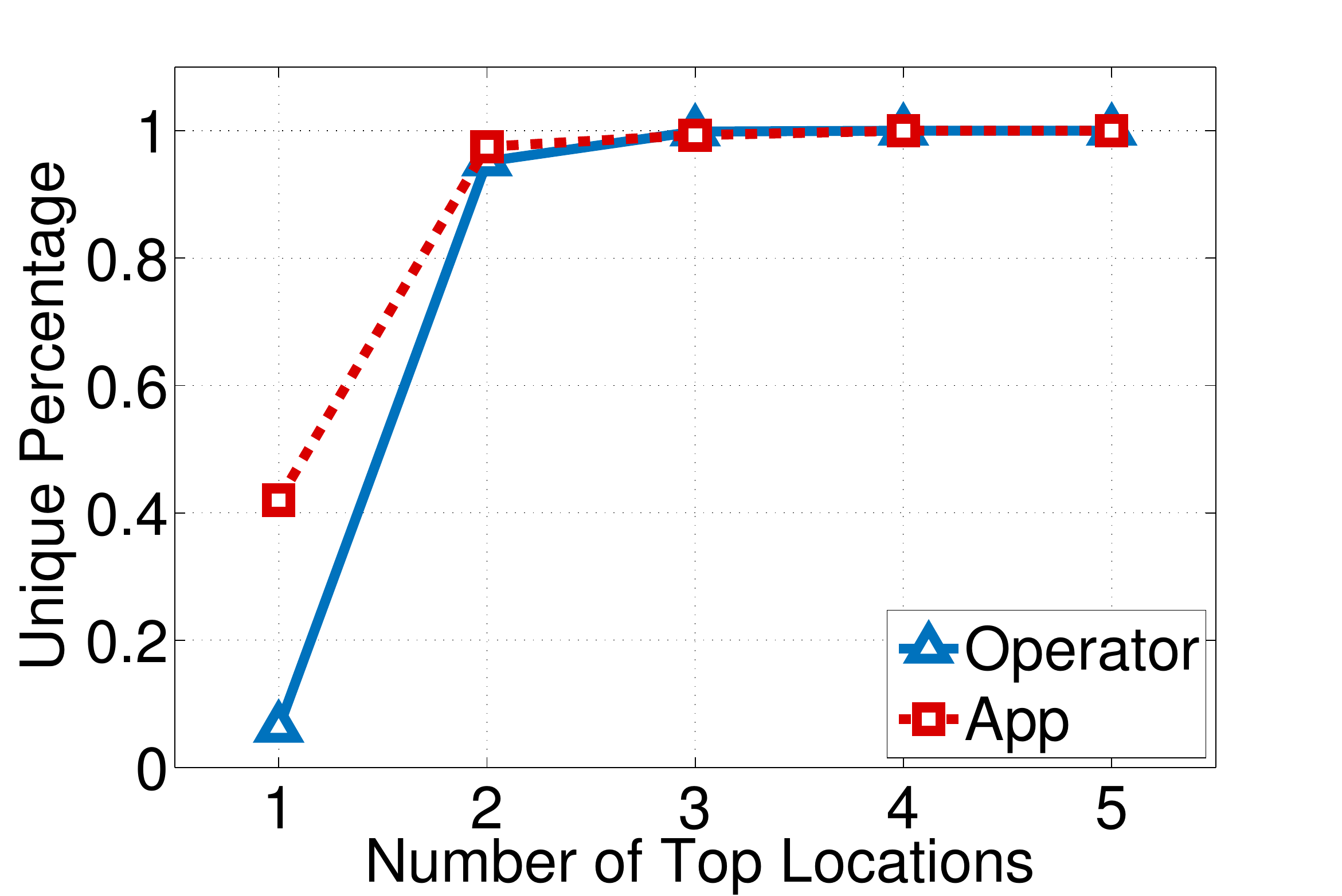}}
\caption{Accuracy, recovery error and uniqueness of the recovered trajectories, where \#1, \#2 and \#3 represent the recovered results after step 1, step 2 and step 3, respectively.} \label{fig:purity}
\end{figure*}

On the other hand, to quantify how far do the wrongly recovered spatiotemporal points deviate from the ground truth, we define the recovery error of each recovered spatiotemporal point as the spatial distance between each recovered spatiotemporal points and the original points in the ground truth. Formally, we compute the recovery error of recovered spatiotemporal point $z^t_i$ as

\begin{equation}
E^t_i =  distance(z^t_i, y^t_i),
\end{equation}\\
where $distance(z^t_i, y^t_i)$ is the euclidean distance between $z^t_i$ and $y^t_i$.

Last but not least, we evaluate the uniqueness of the recovered trajectories to quantify the possibility of linking the recovered trajectories with victims by re-identification. For example, if the attacker knows the locations of victims' homes and working places, the uniqueness measures the possibility that he can uniquely distinguish victims' recovered trajectories. Therefore, we define the uniqueness as the percentage of recovered trajectories that can be uniquely distinguished by their most frequent $k$ locations(Top-$K$), where the Top-$2$ locations usually corresponding to home and working places.

\subsection{Attack Performance}

We quantify the privacy leakage in aggregated mobility data by testing the designed attack system, and evaluating the performance in Figure~\ref{fig:purity}, where \#1, \#2 and \#3 represent the recovered results after step 1, step 2 and step 3, respectively. From Figure~\ref{fig:purity}(a), we observe that in both datasets the recovery accuracy remains high during each step of our attack system. For the application dataset, the accuracy reaches up to 98\% in the first step and slowly decreases to 91\% in the final step. It indicates that our attack system can correctly recovered 98\% of nighttime trajectories and 91\% of whole trajectories. On the other hand, the accuracy for operator data slowly decreases from 95\% to 73\%, which indicates that our system recovers most of the trajectories correctly. These results demonstrate that our system can accurately recover most of the spatiotemporal points in each trajectory, which achieves effective attack on the aggregated mobility data.

From the perspective of recovery error, we present the CDF of the final recovered trajectories' error in Figure~\ref{fig:purity}(b). The results show that only 21\% and 8\% of the recovered spatiotemporal points have a recovery error more than 1,000 meters in mobile operator dataset and application dataset, respectively. In addition, 6\% and 1\% of the recovered spatiotemporal points deviate from the ground truth at distance between 0$\sim$1,000 meters in operator and application datasets, respectively. These results indicate that our attack system is able to recover trajectories with small deviation. On the other hand, we present the percentage of the unique recovered trajectories given the Top-$K$ locations in Figure~\ref{fig:purity}(c). From the results, we can observe that given the two most frequent locations of the recovered trajectories, over 95\% of them can be uniquely distinguished. Therefore, the results indicate that the recovered trajectories are very unique and vulnerable to be reidentified with little external information.

To conclude, the above evaluations indicate that our attack system is effective in breaching the privacy of aggregated mobility data in terms of recovering mobile users' trajectories with high recovery accuracy and low recovery error. In addition, the recovered trajectories have a high possibility to be linked to the victims with external information provided, such as two most frequent locations. These results suggest that the privacy leakage is surprisingly severe even in publishing aggregated mobility data, which contradicts the conventional wisdom and appeals attention to investigate privacy problem in such dataset.

\subsection{Impact of Factors}

After demonstrating the severe privacy leakage in aggregated mobility data, we study the main factors influencing the privacy preservation. Intuitively, datasets consist of large-scale and low spatiotemporal resolution trajectories are often considered to be safer. Therefore, we carry out experiments to quantify the influence of spatial resolution, temporal resolution and dataset scale on the privacy preservation of publishing aggregated mobility data on both dataset. To avoid redundant, we only demonstrate the results of mobile operator dataset since the application dataset yields similar results.

\begin{figure}[t]
\centering
\subfigure[Accuracy]{\includegraphics[width=0.235\textwidth]{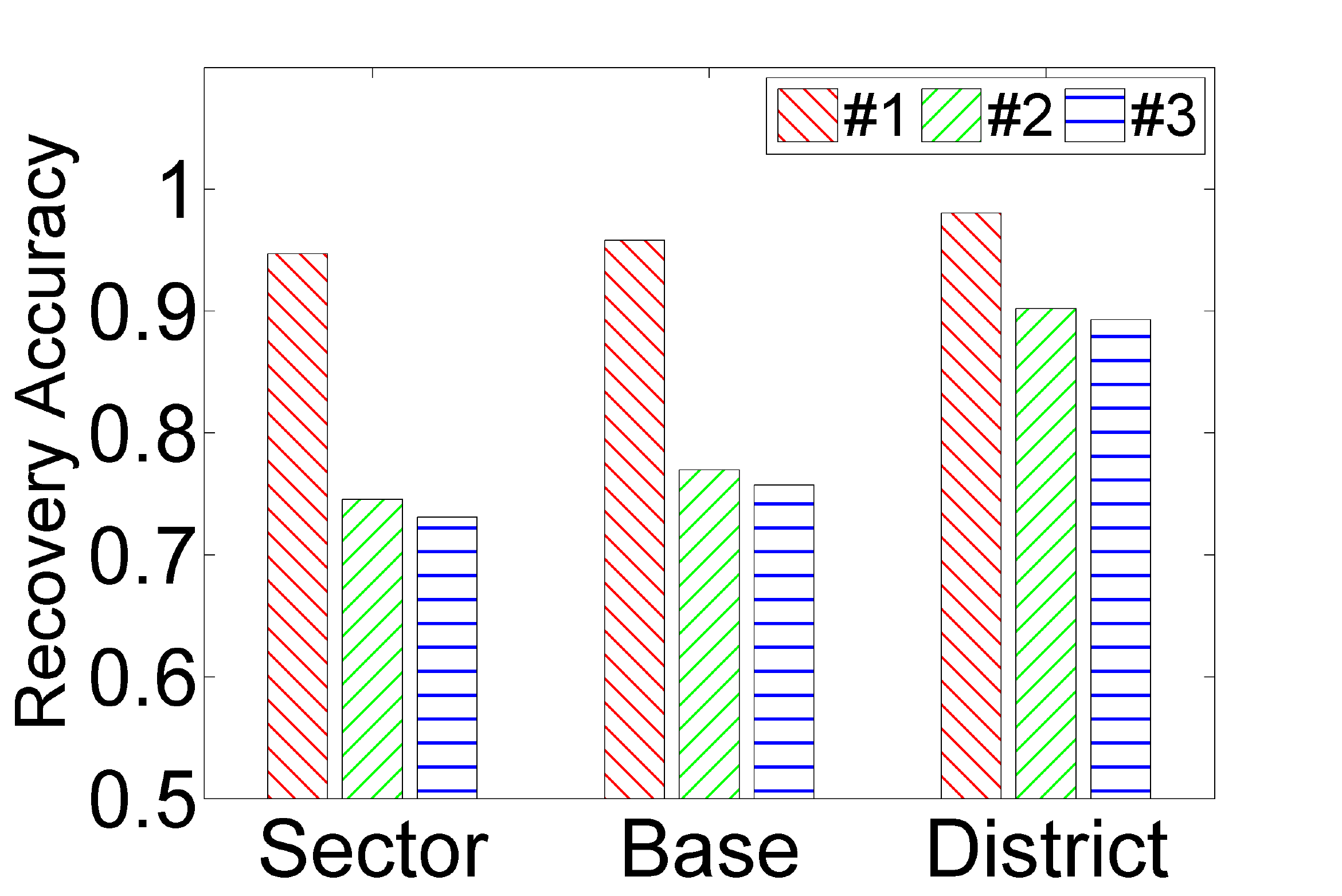}}
\subfigure[Uniqueness]{\includegraphics[width=0.235\textwidth]{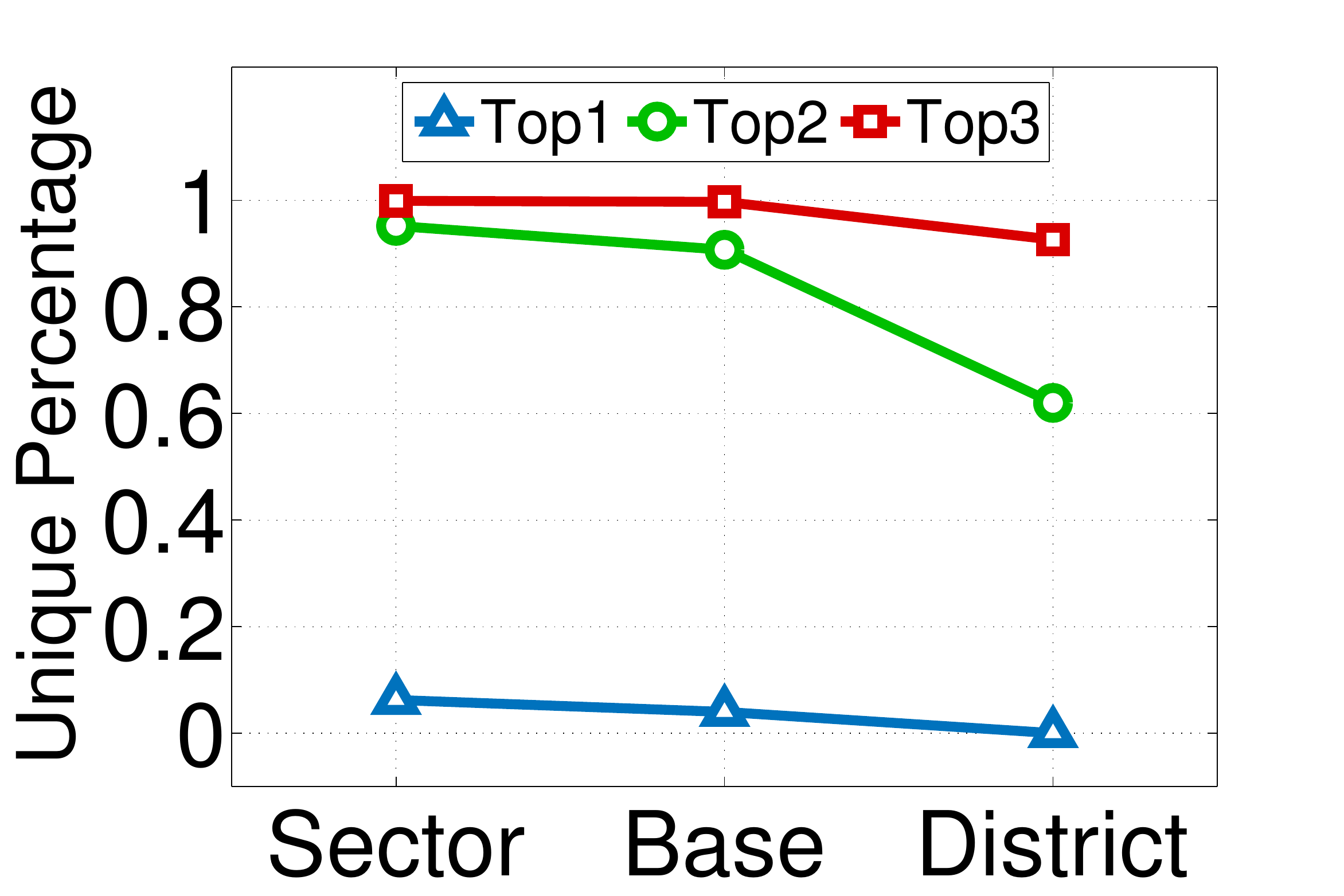}}
\caption{The impact of spatial resolution on privacy leakage. } \label{fig:spatial}
\end{figure}

\textbf{Spatial Resolution.} We first look at how does the spatial resolution influence the privacy leakage. The initial spatial resolution of our investigated datasets is sector of base stations. We aggregate the spatiotemporal points into the resolution of base station and administration district, where each base station has 2 or 3 sectors and the administration district typically covers about 100 base stations. Then, we evaluate the performance of the attack system, and present the recovery accuracy in Figure~\ref{fig:spatial}(a). We can observe that the recovery accuracy increases as the spatial resolution decreases, which indicates more severe privacy leakage in spatial coarse-grained datasets. More specifically, when the spatial resolution reduces from sectors to administration districts, the recovery accuracy increases from 73\% to 89\%, which contradicts with our intuition. On the other hand, Figure~\ref{fig:spatial}(b) demonstrates that the uniqueness of recovered trajectories decreases as the spatial resolution decreases. To be precise, the percentage of uniquely distinguished rate with Top-$2$ locations provided decreases from 95\% to 62\% as the spatial resolution reduces from sectors to administration districts. It suggests that the possibility of linking mobile users and recovered trajectories decreases as spatial resolution decreases. The underlying reason is probably that human mobility is more predictable but less unique in spatial coarse-grained datasets.

\begin{figure}[t]
\centering
\subfigure[Accuracy]{\includegraphics[width=0.235\textwidth]{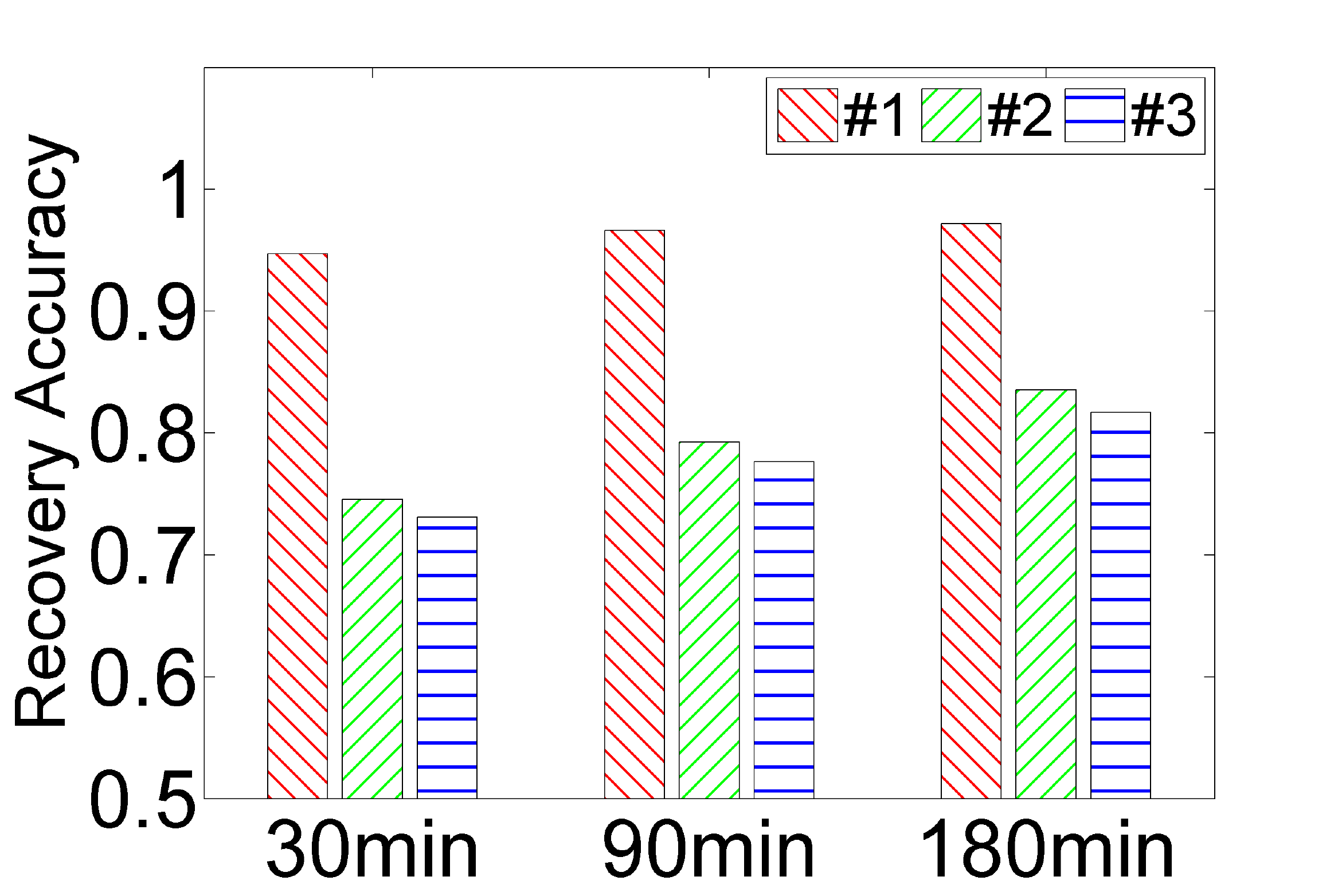}}
\subfigure[Uniqueness]{\includegraphics[width=0.235\textwidth]{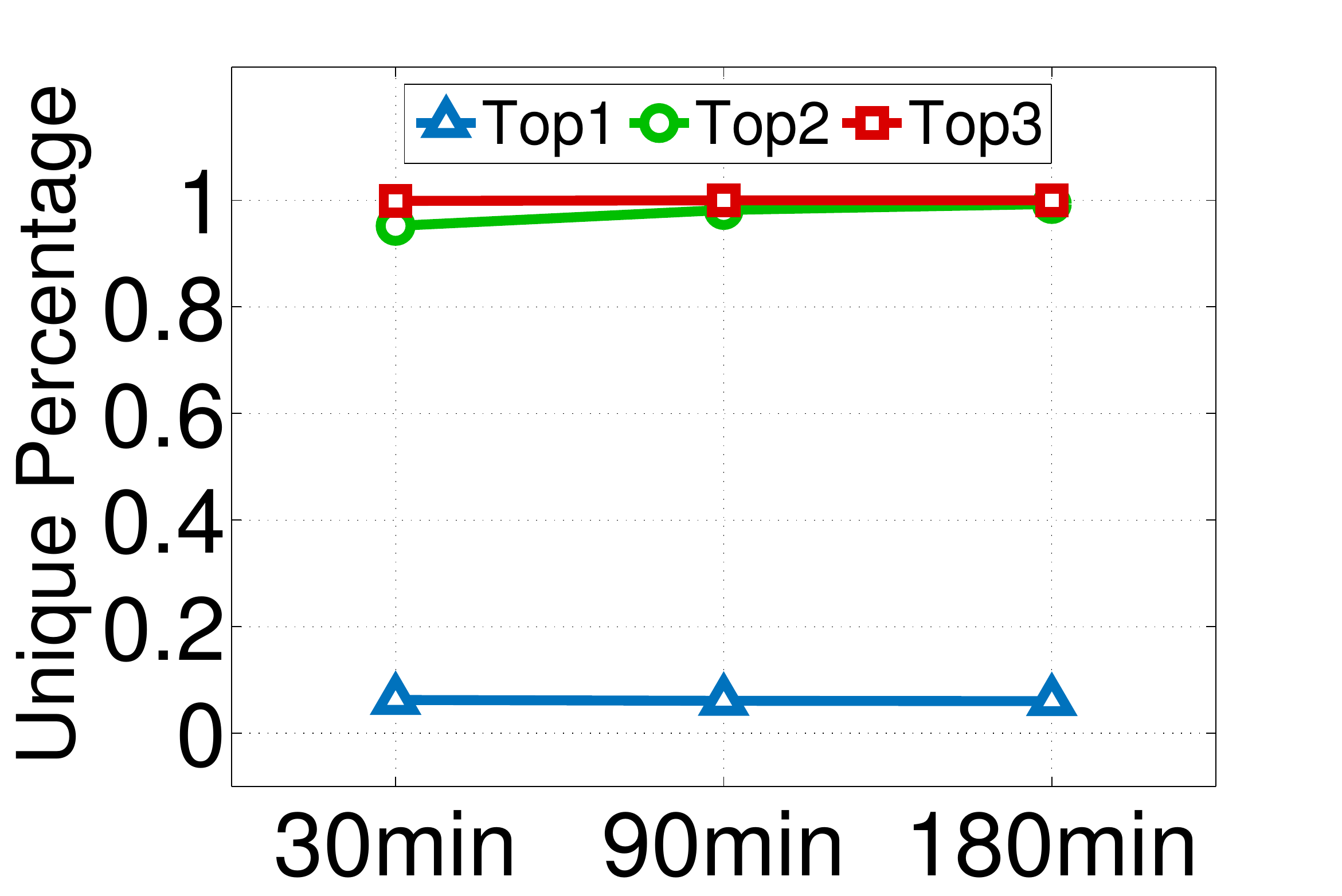}}
\caption{The impact of temporal resolution on privacy leakage. } \label{fig:temporal}
\end{figure}

\begin{figure}[t]
\centering
\subfigure[Accuracy]{\includegraphics[width=0.235\textwidth]{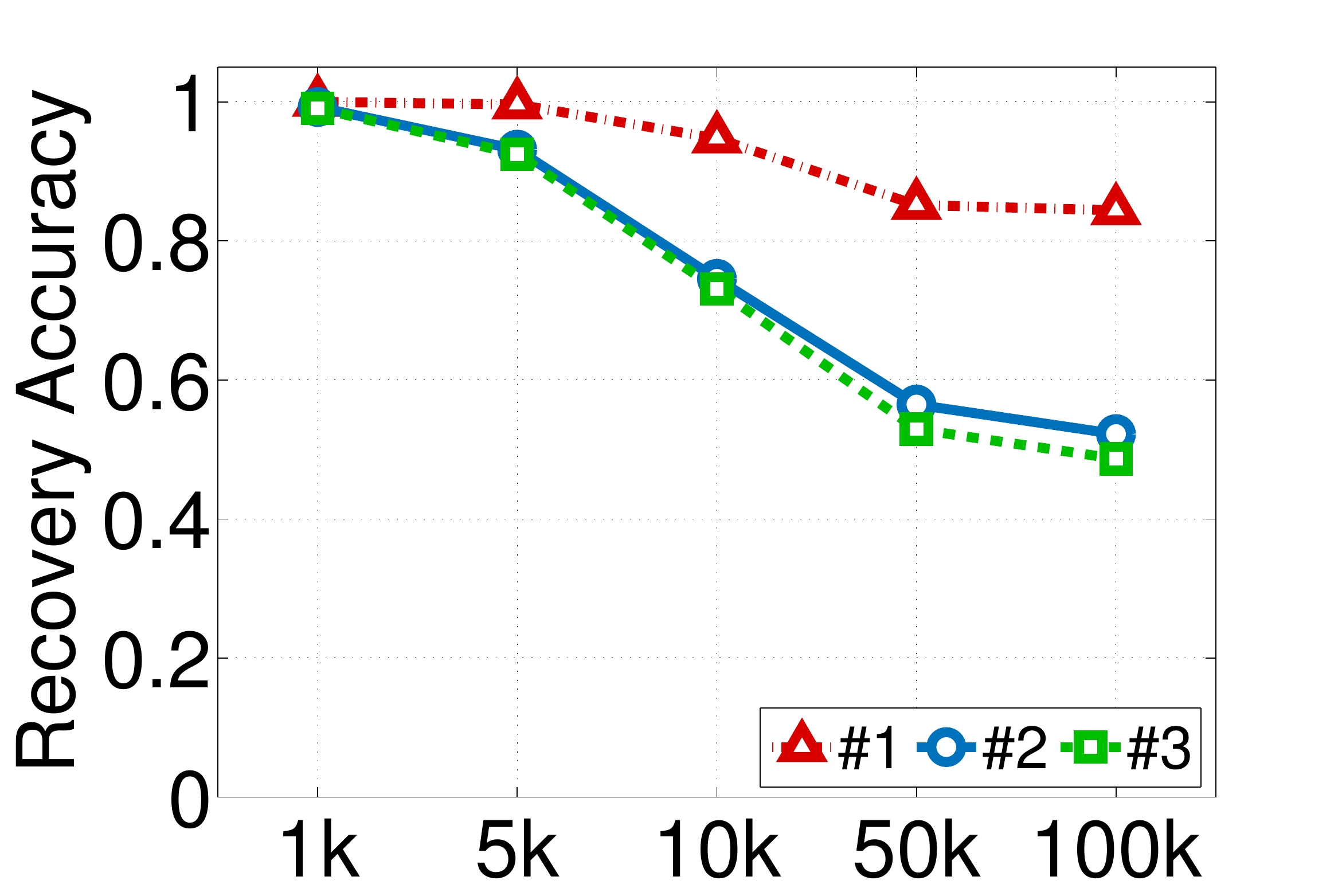}}
\subfigure[Uniqueness]{\includegraphics[width=0.235\textwidth]{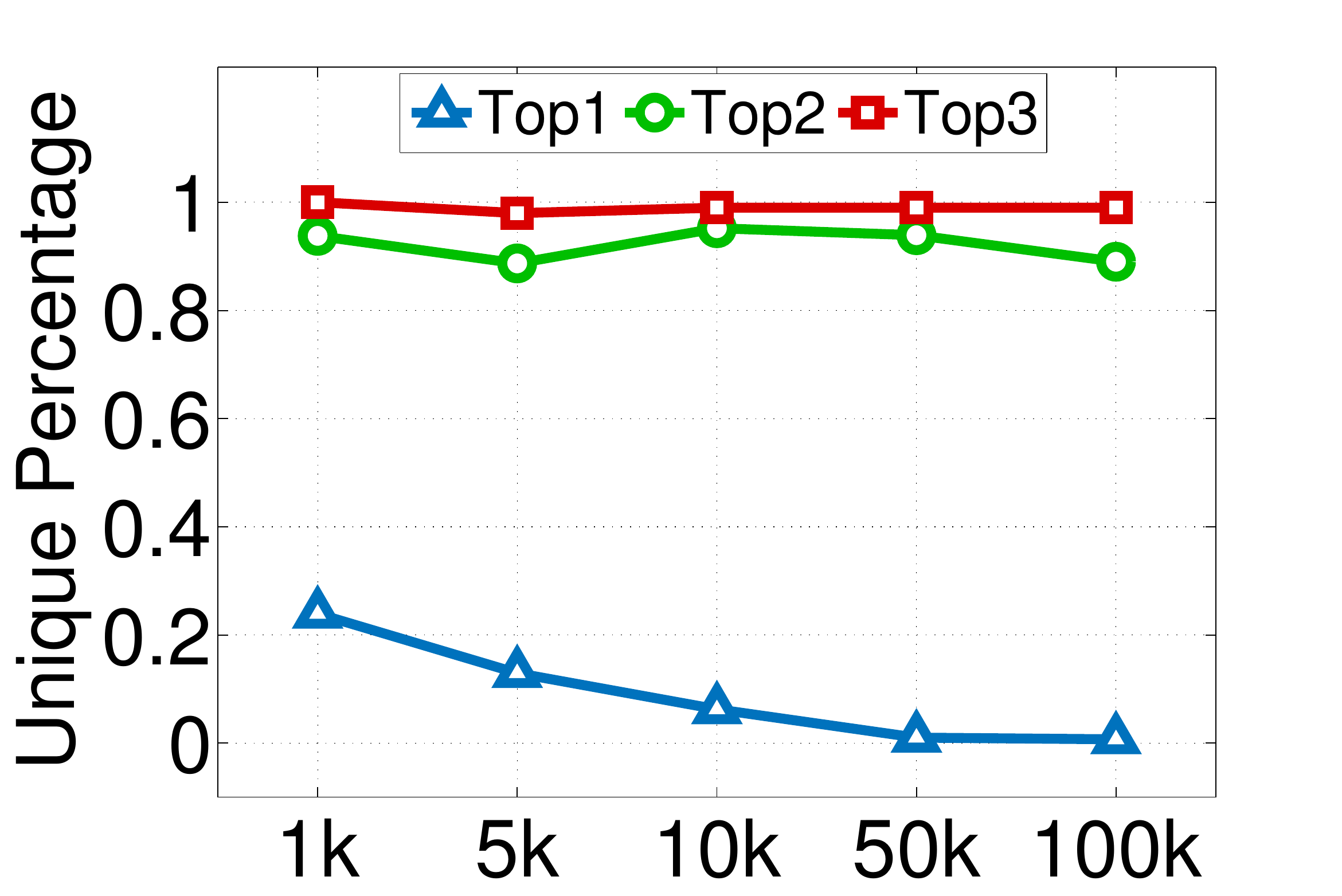}}
\caption{The impact of number of trajectories on privacy leakage. } \label{fig:user}
\end{figure}

\textbf{Temporal resolution.} Temporal resolution is another important factors for mobility datasets. To study its impact, we first produce aggregated mobility data with different temporal resolution, i.e., 30 minutes, 90 minutes and 180 minutes, and then evaluate the performance of the attack system showing in Figure~\ref{fig:temporal}, where \#1, \#2 and \#3 represent the recovered results after step 1, step 2 and step 3, respectively. Observing Figure~\ref{fig:temporal}(a), we surprisingly find out that as the temporal resolution decreases from 30 to 180 minutes, the accuracy of final recovered trajectories slightly increases from 73\% to 82\%. It is probably because when the temporal resolution is low, each trajectory contains fewer spatiotemporal points, which capture more frequent and regular behavior of mobile users in the aggregation process, and hence are more predictable. On the other hand, Figure~\ref{fig:temporal}(b) reveals that the uniqueness of recovered trajectories does not decreases as temporal resolution decreases. More importantly, with Top-$2$ locations provided the uniquely distinguished rate even slightly increases from 95\% to 98\%, which is contrary to decreasing spatial resolution. The underlying reason is probably because mobile users' trajectories are more unique in spatial domain, and hence reducing temporal resolution does not help to make them less unique. As a result, the accuracy and uniqueness of recovered trajectories do not decrease with temporal resolution as expected, which indicates that the attack system is robust under different temporal resolutions.

\textbf{Number of Users.} Since our attack system exploits the uniqueness of individual's trajectory, the number of individuals is assumed to be an important factor that influences the privacy breach. To evaluate this assumption, we randomly sample subsets of different number of individuals from the investigated datasets, and then utilize the proposed attack system to recover the trajectories from each sampled subset. The obtained results are presented in Figure~\ref{fig:user}, where \#1, \#2 and \#3 represent the recovered results after step 1, step 2 and step 3, respectively. Observing Figure~\ref{fig:user}(a), we find out that when the dataset contains only 1,000 mobile users' trajectories, the accuracy is around 0.99, which indicates that the attack system correctly recovers 99\% of spatiotemporal points. In addition, the recovery accuracy indeed decreases as scale of dataset increases. However, when the number of users reaches to 100,000, the attack system is still able to recover the trajectories at accuracy around 50\%. On the other hand, Figure~\ref{fig:user}(b) shows that the uniqueness of recovered trajectories decreases as the scale of datasets grows when only Top-$1$ location is provided. However, with more external information provided, such as Top-$2$ or Top-$3$ locations, the uniquely distinguished rate is stable and remains above 85\% with datasets of different scale. These results reveal that increasing the size of datasets can indeed reduce the accuracy of recovered trajectories, but it cannot prevent attackers link the recovered trajectories with mobile users when more than Top-$2$ locations are provided. More importantly, the evaluation demonstrate that the attack system is valid at the scale of tens of thousands to hundreds of thousands individuals in terms of correctly recovering most of the trajectories.


\section{Discussion}

Through designing and evaluating an attack system for aggregated mobility data, we reveal that simply aggregating mobile user's mobility traces does not provide much privacy preservation as one might expect. Now, we intend to discuss the potential solutions to mitigate such privacy risk, and explore the privacy problem in general statistic data.

\subsection{Potential Mitigation Solutions}
An effective attack system for aggregated mobility data is required to meet two conditions: first, the recovered trajectory is accurate; second, the recovered trajectory can be linked with victim when certain external information is provided. Therefore, the mitigation solutions should be designed to prevent the attackers from meeting both of the conditions. We discuss two potential mitigation solutions --- \emph{generalization} and \emph{perturbation}. 


\textbf{Generalization:} Generalization is a widely adopted scheme that reduces the spatiotemporal resolution to preserve mobile users' privacy in releasing mobility datasets\cite{zang2011anonymization,gramaglia2015hiding}.
Reducing spatiotemporal resolution can potentially reduce the uniqueness of recovered trajectories, because different users' trajectories are more similar when they are coarse-grained. Our previous experiments reveal that spatial generalization can indeed reduce the uniqueness of recovered trajectories, which prevents the attackers from linking them with the victims. In addition, coarse-grained trajectories are also less sensitive because they do not give away the exact locations of mobile users. Therefore, generalization is a valid privacy preserving scheme in releasing aggregated mobility data.


\textbf{Perturbation:} Perturbation is another popular privacy preserving solution in releasing mobility data, which protects individual's privacy by adding noise to the original data \cite{acs2014case, abul2010anonymization}. Since the attack system mainly bases on the behaviours of human mobility, the perturbation scheme should be designed to prevent the attackers from exploiting them. For example, noise can be added to the mobility records at each mobile user's most frequent locations to make their trajectories less regular. In addition, it can also make mobile users' trajectories less unique by eliminating several sensitive mobility records. In short, a well designed perturbation scheme can reduce the regularity and uniqueness of mobile users' trajectories, which has the potential for preserving mobile users' privacy in aggregated mobility data.



\subsection{Privacy Problem in General Statistic Data}
Our paper reveals the potential privacy breach in aggregated mobility data, which is a kind of statistic information of mobile users. Therefore, the follow up question will be: is this privacy problem universal in releasing statistic data? Or, to put it another way, can individual's information be inferred from general statistic data? Inspired by our study, we notice that two key features of the released data facilitate the attack: a)there are patterns in the records created by same individuals --- \emph{regularity}. b) those patterns are different across different individuals --- \emph{uniqueness}. Therefore, the statistic data with similar features is likely suffering from the same privacy breach. Unfortunately, these two features are quite common in the traces left by human, which has been reported in numerous scenarios, such as credit card records \cite{de2015unique}, mobile application usages \cite{welke2016differentiating}, and even web browsing \cite{olejnik2012johnny}. Hence, such privacy problem in statistic data is potentially severe and universal, which calls for immediate attentions from both academy and industry. Evaluating the privacy breaches in different scenarios and developing a generic privacy model for releasing statistic data are left for future works.

\section{Related Work}

Human mobility behaviors and patterns have been studied via mobile cellular data for a decade. By analyzing operators recorded mobility dataset, on one hand, individual mobility is observed with a high degree of temporal and spatial regularity\cite{gonzalez2008understanding}. Such regularity can be potentially predicted with probability of 93\% \cite{song2010limits} and statistically modeled with high accuracy\cite{song2010modelling}. On the other hand, individual mobility is revealed with high uniqueness, i.e., with only several most frequently visited locations\cite{zang2011anonymization} or several random spatiotemporal points\cite{de2013unique, de2015unique}, an individual can be uniquely identified. Instead of studying the regularity patterns or unique mobility behaviors, in this paper we utilize these two fundamental human mobility laws to investigate the privacy issues, i.e., achieving user trajectory recovery from the aggregated mobility datasets.

Despite the increasing privacy concerns about daily Internet usage and web accessing\cite{liu2011analyzing, krishnamurthy2006generating,le2011know, liu2015learning}, recent studies have demonstrated that the privacy risk of releasing anonymized human mobility datasets is severe, since a very high percentage of individuals can be re-identified with their spatial (only locations) \cite{zang2011anonymization,gramaglia2015hiding,kido2005protection, monreale2010movement, sui2016iwqos} or spatiotemporal (with both locations and time information) records\cite{de2013unique}. As a result, a number of privacy-preserving techniques have been proposed with the method of identifier replacement\cite{song2014not}, generalization\cite{gramaglia2015hiding,monreale2010movement}, suppression\cite{garfinkel2006privacy}, or perturbations and permutations\cite{abul2010anonymization, domingo2012microaggregation}, to make sure the released datasets comply with the guideline of $k$-$anonymity$\cite{sweeney2002k, gramaglia2015hiding} and $l$-$diversity$ \cite{machanavajjhala2007diversity, sui2016iwqos}. These privacy techniques are designed under the assumption that unique identifiers of individuals, replaced with random sequence or encrypted, are kept in the released datasets. Therefore, it is completely different with our privacy problem in only publishing aggregated mobility data, where the mobility records belonging to same individuals cannot be associated together.

Recent works also consider the privacy problem in releasing aggregated statistics. However, previous works mainly focused on protecting the membership information of individuals. Acs et al.\cite{acs2014case} implemented a differential privacy scheme on aggregated population density data, which provided provable privacy guarantee that the adversaries cannot determine whether a given mobility record is in the dataset or not. In addition, Dwork et al.\cite{dwork2015robust} proposed a privacy framework that prevented the adversaries to infer the membership information of an individual given the statistic information of a DNA dataset. Different from previous works, our paper targets at a more general privacy problem on inferring personal information, not just membership information, from the aggregated statistic of a dataset.


\section{Conclusion}

In this paper, we identify and evaluate the risks of trajectory recovery attack in the aggregated mobility dataset. To the best of our knowledge, we are the first to recognize and study the privacy problem of inferring individual's information from statistic data. Our investigation reveals that there is serious privacy leakage in the aggregated mobility data since individuals' trajectories can be recovered with high accuracy.
In addition, our evaluation demonstrates that the spatiotemporal resolution and scale of datasets have notable impact on the privacy breach.
We believe that this work opens a new angle of protecting the privacy in publishing and sharing statistic data, which paves the way to more advanced privacy preserving mechanisms.

\section{Acknowledgments}
This work is supported by National Basic Research Program of China (973 Program) (No. 2013CB329105), National Nature Science Foundation of China (No. 61301080), Research Fund of Tsinghua University (No. 20161080099), and Research Fund of Tsinghua-Tencent Join Laboratory.

\newpage
\bibliographystyle{unsrt}
\bibliography{cellular}

\begin{thebibliography}{10}

\bibitem{wang2015data}
R.~Wang, M.~Xue, K.~Liu, et~al.
\newblock Data-driven privacy analytics: A wechat case study in location-based
  social networks.
\newblock In {\em Wireless Algorithms, Systems, and Applications}. Springer,
  2015.

\bibitem{apple2015polilcy}
Apple's commitment to your privacy.
\newblock {\em http://www.apple.com/privacy/}.

\bibitem{blondel2012data}
V.~D. Blondel, M.~Esch, C.~Chan, et~al.
\newblock Data for development: the d4d challenge on mobile phone data.
\newblock {\em arXiv preprint arXiv:1210.0137}, 2012.

\bibitem{acs2014case}
G.~Acs and C.~Castelluccia.
\newblock A case study: privacy preserving release of spatio-temporal density
  in paris.
\newblock In {\em ACM SIGKDD}. ACM, 2014.

\bibitem{dengta}
China telcome' big data products.
\newblock {\em http://www.dtbig.com/}.

\bibitem{song2010limits}
C.~Song, Z.~Qu, and N.~Blumm.
\newblock Limits of predictability in human mobility.
\newblock {\em Science}, 2010.

\bibitem{isaacman2011ranges}
S.~Isaacman, R.~Becker, R.~C{\'a}ceres, et~al.
\newblock Ranges of human mobility in \protect{Los Angeles and New York}.
\newblock In {\em IEEE PERCOM Workshops}. IEEE, 2011.

\bibitem{isaacman2012human}
S.~Isaacman, R.~Becker, R.~C{\'a}ceres, et~al.
\newblock Human mobility modeling at metropolitan scales.
\newblock In {\em ACM MOBISYS}. ACM, 2012.

\bibitem{seshadri2008mobile}
M.~Seshadri, S.~Machiraju, A.~Sridharan, et~al.
\newblock Mobile call graphs: beyond power-law and lognormal distributions.
\newblock In {\em ACM KDD}. ACM, 2008.

\bibitem{wang2013inferring}
Y.~Wang, H.~Zang, and M.~Faloutsos.
\newblock Inferring cellular user demographic information using homophily on
  call graphs.
\newblock In {\em IEEE INFOCOM WKSHPS}. IEEE, 2013.

\bibitem{wesolowski2012quantifying}
A.~Wesolowski, N.~Eagle, A.~J. Tatem, et~al.
\newblock Quantifying the impact of human mobility on malaria.
\newblock {\em Science}, 2012.

\bibitem{saravanan2013exploring}
M.~Saravanan, P.~Karthikeyan, and A.~Aarthi.
\newblock Exploring community structure to understand disease spread and
  control using mobile call detail records.
\newblock {\em NetMob D4D Challenge}, 2013.

\bibitem{douglass2015high}
R.~W. Douglass, D.~A. Meyer, M.~Ram, et~al.
\newblock High resolution population estimates from telecommunications data.
\newblock {\em EPJ Data Science}, 2015.

\bibitem{wang2015understanding}
H.~Wang, F.~Xu, Y.~Li, et~al.
\newblock Understanding mobile traffic patterns of large scale cellular towers
  in urban environment.
\newblock In {\em ACM IMC}. ACM, 2015.

\bibitem{sweeney2002k}
L.~Sweeney.
\newblock k-anonymity: A model for protecting privacy.
\newblock {\em International Journal of Uncertainty, Fuzziness and
  Knowledge-Based Systems}, 2002.

\bibitem{de2015unique}
Y.~de~Montjoye, L.~Radaelli, V.~K. Singh, et~al.
\newblock Unique in the shopping mall: On the reidentifiability of credit card
  metadata.
\newblock {\em Science}, 2015.

\bibitem{zang2011anonymization}
H.~Zang and J.~Bolot.
\newblock Anonymization of location data does not work: A large-scale
  measurement study.
\newblock In {\em ACM Mobicom}. ACM, 2011.

\bibitem{gramaglia2015hiding}
M.~Gramaglia and M.~Fiore.
\newblock Hiding mobile traffic fingerprints with glove.
\newblock {\em ACM CoNEXT}, 2015.

\bibitem{barabasi2005origin}
A.-L. Barabasi.
\newblock The origin of bursts and heavy tails in human dynamics.
\newblock {\em Nature}, 2005.

\bibitem{machanavajjhala2007diversity}
A.~Machanavajjhala, D.~Kifer, J.~Gehrke, et~al.
\newblock l-diversity: Privacy beyond k-anonymity.
\newblock {\em ACM TKDD}, 2007.

\bibitem{de2013unique}
Y.~de~Montjoye, C.~A. Hidalgo, M.~Verleysen, et~al.
\newblock Unique in the crowd: The privacy bounds of human mobility.
\newblock {\em Scientific reports}, 2013.

\bibitem{dantzig1998linear}
G.~B. Dantzig.
\newblock {\em Linear programming and extensions}.
\newblock Princeton university press, 1998.

\bibitem{kuhn1955hungarian}
H.~W. Kuhn.
\newblock The \protect{Hungarian} method for the assignment problem.
\newblock {\em Naval research logistics quarterly}, 1955.

\bibitem{abul2010anonymization}
O.~Abul, F.~Bonchi, and M.~Nanni.
\newblock Anonymization of moving objects databases by clustering and
  perturbation.
\newblock {\em Information Systems}, 2010.

\bibitem{welke2016differentiating}
Pascal Welke, Ionut Andone, Konrad Blaszkiewicz, and Alexander Markowetz.
\newblock Differentiating smartphone users by app usage.
\newblock In {\em Proceedings of the 2016 ACM International Joint Conference on
  Pervasive and Ubiquitous Computing}, pages 519--523. ACM, 2016.

\bibitem{olejnik2012johnny}
Lukasz Olejnik, Claude Castelluccia, and Artur Janc.
\newblock Why johnny can't browse in peace: On the uniqueness of web browsing
  history patterns.
\newblock In {\em 5th Workshop on Hot Topics in Privacy Enhancing Technologies
  (HotPETs 2012)}, 2012.

\bibitem{gonzalez2008understanding}
M.~C. Gonzalez, C.~A. Hidalgo, and A.-L. Barabasi.
\newblock Understanding individual human mobility patterns.
\newblock {\em Nature}, 2008.

\bibitem{song2010modelling}
C.~Song, T.~Koren, P.~Wang, et~al.
\newblock Modelling the scaling properties of human mobility.
\newblock {\em Nature Physics}, 2010.

\bibitem{liu2011analyzing}
Y.~Liu, K.~P. Gummadi, B.~Krishnamurthy, et~al.
\newblock Analyzing facebook privacy settings: user expectations vs. reality.
\newblock In {\em ACM IMC}. ACM, 2011.

\bibitem{krishnamurthy2006generating}
B.~Krishnamurthy and C.~E. Wills.
\newblock Generating a privacy footprint on the internet.
\newblock In {\em ACM IMC}. ACM, 2006.

\bibitem{le2011know}
S.~Le~B., C.~Zhang, A.~Legout, et~al.
\newblock I know where you are and what you are sharing: exploiting p2p
  communications to invade users' privacy.
\newblock In {\em ACM IMC}. ACM, 2011.

\bibitem{liu2015learning}
S.~Liu, I.~Foster, S.~Savage, et~al.
\newblock Who is. com? learning to parse whois records.
\newblock In {\em ACM IMC}. ACM, 2015.

\bibitem{kido2005protection}
H.~Kido, Y.~Yanagisawa, and T.~Satoh.
\newblock Protection of location privacy using dummies for location-based
  services.
\newblock In {\em IEEE ICDEW}. IEEE, 2005.

\bibitem{monreale2010movement}
A.~Monreale, G.~L. Andrienko, N.~V. Andrienko, et~al.
\newblock Movement data anonymity through generalization.
\newblock {\em Transactions on Data Privacy}, 2010.

\bibitem{sui2016iwqos}
K.~Sui, Y.~Zhao, D.~Liu, et~al.
\newblock Your trajectory privacy can be breached even if you walk in groups.
\newblock {\em IEEE/ACM IWQoS}, 2016.

\bibitem{song2014not}
Y.~Song, D.~Dahlmeier, and S.~Bressan.
\newblock Not so unique in the crowd: a simple and effective algorithm for
  anonymizing location data.
\newblock In {\em PIR@ SIGIR}, 2014.

\bibitem{garfinkel2006privacy}
S.~Garfinkel.
\newblock Privacy protection and \protect{RFID}.
\newblock In {\em Ubiquitous and Pervasive Commerce}. Springer, 2006.

\bibitem{domingo2012microaggregation}
J.~Domingo-Ferrer and R.~Trujillo-Rasua.
\newblock Microaggregation-and permutation-based anonymization of movement
  data.
\newblock {\em Information Sciences}, 2012.

\bibitem{dwork2015robust}
Cynthia Dwork, Adam Smith, Thomas Steinke, Jonathan Ullman, and Salil Vadhan.
\newblock Robust traceability from trace amounts.
\newblock In {\em Foundations of Computer Science (FOCS), 2015 IEEE 56th Annual
  Symposium on}, pages 650--669. IEEE, 2015.

\end{thebibliography}

\end{document}